\documentclass[%
 reprint,
 amsmath,amssymb,
 aps,soul,
pre,
]{revtex4-2}

\usepackage{graphicx}
\usepackage{dcolumn}
\usepackage{bm}

\renewcommand{\selectlanguage}[1]{}

\begin{document}

\preprint{APS/ES12411}

\title{A Point-cloud Clustering \& Tracking Algorithm for Radar Interferometry}

\author{Magnus F Ivarsen}
\altaffiliation[Also at ]{Department of Physics and Engineering Physics, University of Saskatchewan, Saskatoon, Canada}
\affiliation{Department of Physics, University of Oslo, Oslo, Norway}%

\author{Jean-Pierre St-Maurice}
\altaffiliation[Also at ]{Department of Physics and Astronomy, University of Western Ontario, London, Ontario, Canada}
\author{Glenn C Hussey}
\affiliation{Department of Physics and Engineering Physics, University of Saskatchewan, Saskatoon, Canada}%
\author{Devin R Huyghebaert}
\affiliation{Department of Physics and Technology, University of Tromsø, Tromsø, Norway}%
\author{Megan D Gilles}
\affiliation{Department of Chemistry and Physics , Mount Royal University, Calgary, Canada}%

\date{\today}

\begin{abstract}
In data mining, density-based clustering, which entails classifying datapoints according to their distributions in some space, is an essential method to extract information from large datasets. With the advent of software-based radio, ionospheric radars are capable of producing unprecedentedly large datasets of plasma turbulence backscatter observations, and new automatic techniques are needed to sift through them. We present an algorithm to automatically identify and track clusters of radar echoes through time, using \texttt{dbscan}, a celebrated density-based clustering method for noisy point-clouds. We demonstrate our algorithm's efficiency by tracking turbulent structures in the E-region ionosphere, the so-called radar aurora. Through conjugate auroral imagery, as well as \emph{in-situ} satellite observations, we demonstrate that the observed turbulent structures generally track the motion of auroras. What is more, the radar aurora bulk motions exhibit key qualities of auroral electric field enhancements that has previously been observed with various instruments. We present preliminary statistical results using our new method, and briefly discuss the method's limitations and potential future adaptations.
\end{abstract}

\maketitle


\section{\label{sec:level1}Introduction}

In the practice of data mining, `data clustering' attempts to segment a dataset into groups or classes that share common traits and characteristics. These characteristics are not immediately discernible in individual datapoints, and their association (but not the \emph{cause} of their association) can be revealed by clustering \cite{birantSTDBSCANAlgorithmClustering2007}. Density-based clustering, in turn, achieves this goal by classifying datapoints based on their density inside a space defined by two or more variables, a point-cloud. In the present study, we apply density-based clustering to interferometry radar data in the field of ionospheric research, an endeavour that will enable key insights into ionospheric physics.

With recent advances in radar and lidar (optical radar) technologies, spatial point-clouds are becoming increasingly common. For example, self-driving cars rely on real-time obstacle-detection using lidar scans, in which fast and accurate point-cloud clustering is vital \cite{mcdanielTerrainClassificationIdentification2012,liRealtimeSelfdrivingCar2017}. Lidar mounted on aircraft are frequently used in geographic surveying \cite{hodgsonEvaluationLidarderivedElevation2005} and for discovering future archaeological sites \cite{schindlingLiDARToolArchaeological2014}. In the survey of Earth's space environment, the \textsc{icebear}~3D radar represents a recent advancement in radar technology that is capable of producing 3D point-clouds of plasma turbulence echoes in the lower ionosphere \cite{lozinskyICEBEAR3DLowElevation2022}. In a recent paper, \cite{ivarsenAlgorithmSeparateIonospheric2023} demonstrated that such echoes tend to be organized in tight clusters, and the degree of clustering is directly reflecting the filamentation of geomagnetic currents in the ionosphere \cite{ivarsenDistributionSmallScaleIrregularities2023,ivarsen_turbulence_embedded_2024}.

Unlike conventional radar and lidar imaging, in which the subject is a tangible man-made or natural object, ionospheric radar images are produced by turbulent, or stochastic, processes \cite{fejerHighLatitudeEregion1987}. Measurements of stochastic processes are unpredictable and must in general be understood in terms of statistical representation, which is usually obtained through Fourier analysis \cite{ochiOceanWavesStochastic1998}. The turbulent shapes that are imaged by the \textsc{icebear} radar are thus viable at any time to bifurcate, merge, dissipate, and grow, making efficient tracking of their bulk motion a challenging endeavour.

\subsection{\label{sec:level2}Plasma Turbulence in the E-region}

The ionosphere, a partially ionized layer of gas existing in Earth's atmosphere at altitudes above $\sim90$~km, can at times be highly turbulent, with large portions of the ionosphere's plasma exhibiting chaotic fluid properties \cite{hubaIonosphericTurbulenceInterchange1985}. Collectively and ultimately, the turbulence serves to dissipate the energy imparted to Earth from the solar wind during geomagnetic storms. Field-aligned currents and precipitating particles pass electric fields that stir the plasma, which, together with the interplanetary magnetic field, create a gigantic convection pattern that covers Earth's high-latitude regions \cite{dungeyInterplanetaryMagneticField1961,cowleyTUTORIALMagnetosphereIonosphereInteractions2000}, as well as direct large amounts of electromagnetic energy into the atmosphere \cite{keilingAssessingGlobalAlfven2019}. The ionosphere's conductivity, its ability to support electrical currents, is central to how these processes unfold, and small-scale turbulence in the ionosphere can cause stochastic changes to this conductivity \cite{wiltbergerEffectsElectrojetTurbulence2017,st-mauriceRevisitingBehaviorERegion2021}.

At high latitudes, on Earth's nightside, the particles that precipitate into the atmosphere eventually collide with neutral gas particles, whereby they excite light emissions,  creating the aurora \cite{prolssAbsorptionDissipationSolar2004a}. The aurora, appearing to observers on the ground as elongated curtains of shifting, or dancing, light, is accompanied by a conspicuous radio signature in the bottomside (E-region) ionosphere, the radar aurora \cite{hysellRadarAurora2015}. Radars, when pointed in the direction perpendicular to Earth's magnetic field lines, can detect an echo of their outgoing signal, echoes that are produced when that signal is scattered off of sharp density gradients in the plasma. With a 50~MHz radar signal, such sharp gradients are 3-meter waves produced by the Farley-Buneman (FB) instability, a modified two-stream instability mechanism \cite{bunemanExcitationFieldAligned1963,farleyPlasmaInstabilityResulting1963}.

The growth of FB waves in the E-region depends on the electric field being sufficiently strong \cite{fejerIonosphericIrregularities1980}. Part of the transmitted radio signals is subject to Bragg-like scattering, allowing for an echo detection by the radar receiver, if the relative fluctuation level in the plasma density irregularity that caused the scattering is sufficiently high  \cite{bowles_field-aligned_1963}. In addition, as alluded to, FB waves mostly grow in directions perpendicular to Earth's magnetic field \cite{fejerIonosphericIrregularities1980,drexlerNewInsightsNonlocal2002}. The \textsc{icebear} radar, located in Saskatchewan, Canada, beams its signal towards the horizon. Here, given regions of favourable conditions, a continuous distribution of FB turbulence will produce a 3D point cloud every second. These echo point-clouds are frequently observed to cluster around auroral arcs  \cite{huyghebaertPropertiesICEBEARERegion2021,ivarsenTurbulenceAuroras2024}, a crucial observation that is valid for the radar aurora in general \cite{hallSpatialRelationshipLarge1990,bahcivanObservationsColocatedOptical2006,hysellComparingVHFCoherent2012}. Since the aurora is an emergent phenomenon that arises through the interaction between the ionosphere, Earth's magnetosphere, and the solar wind \cite{borovskyQuiescentDiscreteAuroral2019}, we can infer key characteristics of the behaviour of echo clusters: their motion must be governed by magnetospheric processes, often arising from the complex interplay between Alfv\'{e}nic and quasi-static considerations \cite{chastonMotionAurorae2010}.


In the present study, we describe a set of novel methods adapted from the flourishing field of data mining that we applied to automatically detect and track dense clusters of coherent backscatter echoes in the \textsc{icebear}~3D dataset. The \textsc{icebear} radar routinely detects multitudes of echoes in excess of $10^4$ per minute, each of which is associated with a 3D spatial location. In space, then, the E-region turbulence regions frequently form, bifurcate, break up, and reappear, all the while exhibiting some form of motion in the plane perpendicular to Earth's field lines. This motion is complicated and composite, and must be described as \emph{apparent} by an observer on Earth. However, as we demonstrate in a related publication, \cite{ivarsen_deriving_2024}, a clear tendency to follow the motion of electric field enhancements is discernible in the trajectories of \textsc{icebear} echo clusters. The present paper is a \emph{method} paper whose intent is to explain, elucidate, and discuss the process of extracting spatial tracks made by distinct echo clusters, as well as to demonstrate a clear and unambiguous tendency in those tracks to follow the motion of auroral arcs.

\section{Methodology \& data}

As already mentioned, our dataset consists of E-region radio echoes from 3-meter size irregularities produced by the FB instability, as detected  by the \textsc{icebear} (Ionospheric Continuous-wave E-region Bistatic Experimental Auroral Radar), a 49.5~MHz coherent scatter radar in Saskatchewan, Canada \cite{huyghebaertICEBEARAlldigitalBistatic2019}. The \textsc{icebear}~3D dataset is the result of a reconfiguration so that echo locations can be determined with a higher spatial and temporal resolution \cite{lozinskyICEBEAR3DLowElevation2022}. In this reconfiguration, the data processing chain produces a full field-of-view radio image for every range-Doppler time bin,
thereby handling some 200,000 such images per second. In each image, a single target radar echo is identified by a single peak in the brightness distribution, at which point its location in latitude, longitude, and altitude is determined from its bearing and heading. This is 
in addition to the target echo's spectral properties, particularly its Doppler shift.

The present work has been motivated by recent results showing that \textsc{icebear}~3D echoes appear to cluster tightly in the plane perpendicular to Earth's magnetic field lines \cite{ivarsenAlgorithmSeparateIonospheric2023}. Building on this, we have applied the \texttt{dbscan} algorithm to automatically detect clusters in two-dimensional point-clouds \cite{esterDensitybasedAlgorithmDiscovering1996,khanDBSCANPresentFuture2014}. To implement the methods, we have tracked clusters through time with a rudimentary test of similarity from one time-step to the next. We sort the data into time-intervals of three seconds, with a one-second cadence. Based on this approach, we were able  to systematically identify and track turbulent regions as they appeared in the radar FOV.

\begin{figure*}
\includegraphics[width=.86\textwidth]{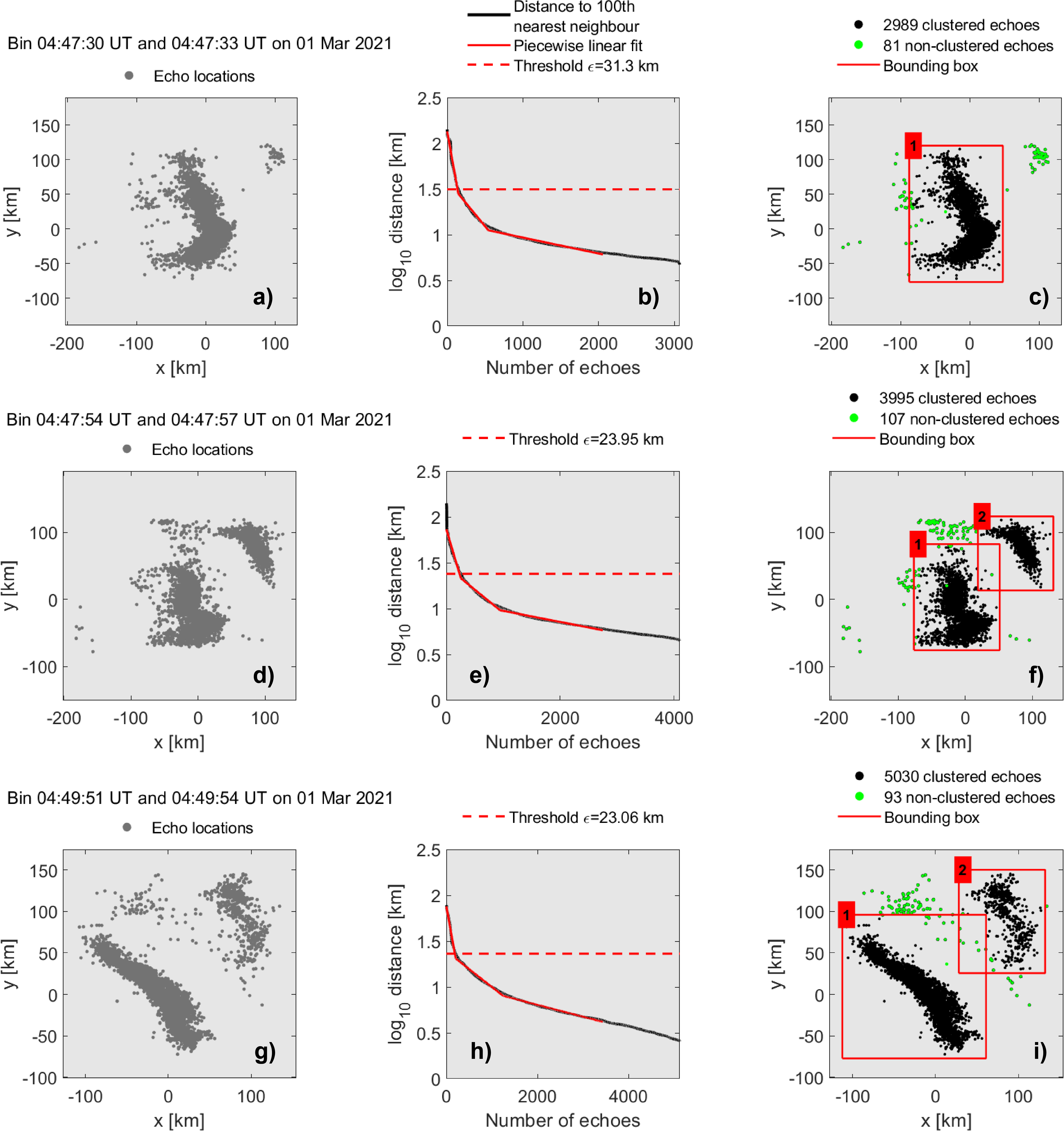}
\caption{\label{fig:ex}Examples of the echo point-cloud clustering categorization  using the \texttt{dbscan} algorithm. Leftmost column: the raw echo locations.  Middle column: $k$-distance plot for all the echoes (with $k=100^\text{th}$ nearest neighbour). A solid red line gives a piecewise linear fit, while a dashed red line shows the determined threshold distance $\epsilon$. Right column: displayed and labeled clusters that were identified. The labels attached to clusters in panels c), f), and i) do not carry over to other panels.}
\end{figure*}

The unsupervised learning algorithm \texttt{dbscan} assigns points to groups, or clusters, essentially by requiring that the $k^\text{th}$ nearest \emph{neighbour} of a point is less than a distance $\epsilon$ away from that point \cite{esterDensitybasedAlgorithmDiscovering1996,sander_density-based_2010}. Points for which this distance is above $\epsilon$ are considered \emph{noise} or outliers, and do not belong to any cluster. The threshold distance $\epsilon$ is determined by a so-called $k$-distance plot, which simply sorts the $k^\text{th}$ nearest neighbour distance metric for each point. The kink or elbow of this curve, where the $k^\text{th}$ nearest neighbour distance metric flattens to a plateau, identifies the characteristic noise threshold $\epsilon$ for a particular point-cloud. Examples of this procedure are shown in Figure~\ref{fig:ex}, where the left column shows \textsc{icebear}~3D point-clouds, the middle column shows the $k^\text{th}$ distance plot associated with each cloud, and the right column shows the resulting clusters that were identified.

In the middle column of Figure~\ref{fig:ex}, we observe that the noise threshold $\epsilon$ takes on various  values, based on the natural tendency for points to cluster in separate point-clouds. Based on extensive testing, we limited the value of $\epsilon$ to be between 3~km and 200~km, and we took the average value of $\epsilon$ determined within a six-second period to determine the transition to a plateau.  
In the right column of Figure~\ref{fig:ex}, we illustrate how, based on our selection criteria,  the \texttt{dbscan} algorithm has identified two clusters in each point-cloud (black data), with a smaller number of non-clustered points (green data).

The red 'bounding boxes' that enclose each cluster of echoes is then used to track echo clusters through time, by requiring that the bounding box size and location of a particular cluster must be reasonably similar to a cluster identified at the previous timestep. The bounding box similarity test consists of minimizing three distance metrics: 1) the longitudinal extent of the bounding box, 2) the latitudinal extent of the bounding box, and 3) the median echo location inside the bounding box. Two consecutive populations that minimize the difference in each of the three distance metrics are determined to be the same population if the metrics are all smaller than a pre-defined distance $D$. For the purpose of identifying and tracking large echo populations through time we have found, through extensive testing, that the procedure favours an abnormally high $D$ (e.g., $D=80$~km),  as well as a relatively high $k$ (e.g., $k=100$). This ensures that echo populations that grow or shrink drastically will still be successfully tracked through time, and echo clusters that consist of very few echoes will be ignored. On the other hand, to accurately determine bulk velocities (which is the main focus of the results of the present paper) the procedure favours relatively low parameters (e.g., $D=80$~km and $k=20$). As we shall demonstrate in the two next sections, this scheme will produce a high number of tracks whose average motion will accurately quantify the bulk motion of the radar aurora. 

\begin{figure}
    \centering
    \includegraphics[width=.35\textwidth]{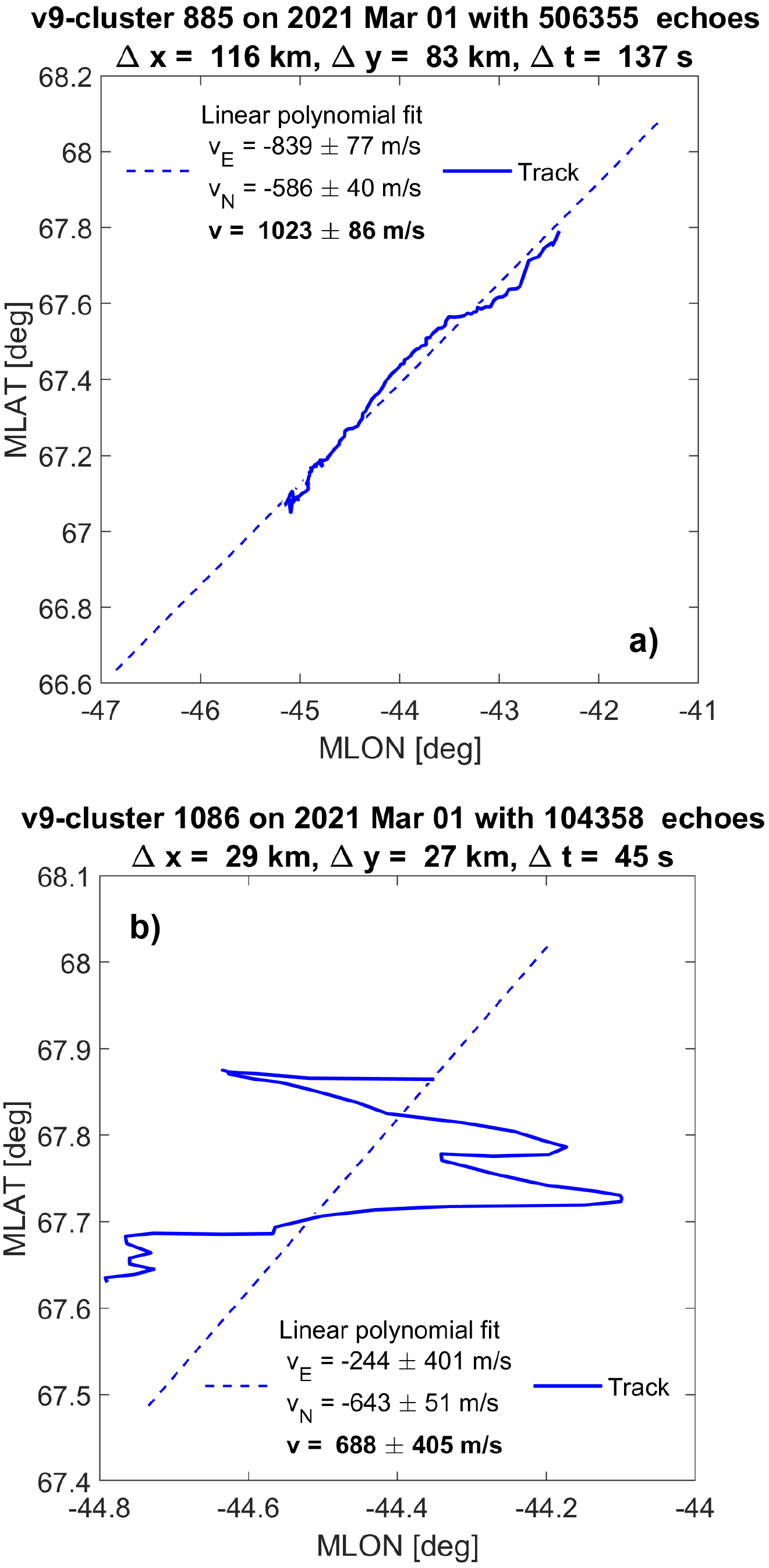}
    \caption{Two example echo cluster tracks (solid blue lines) and their corresponding linear polynomial fits (dashed blue lines). The linear coefficient ($v_E,v_N$) is displayed for both clusters, with the 95-percentile (3-sigma) confidence interval for each fit denoting the error margin.}
    \label{fig:tracks}
\end{figure}

\begin{figure*}
\centering
\includegraphics[width=.9\textwidth]{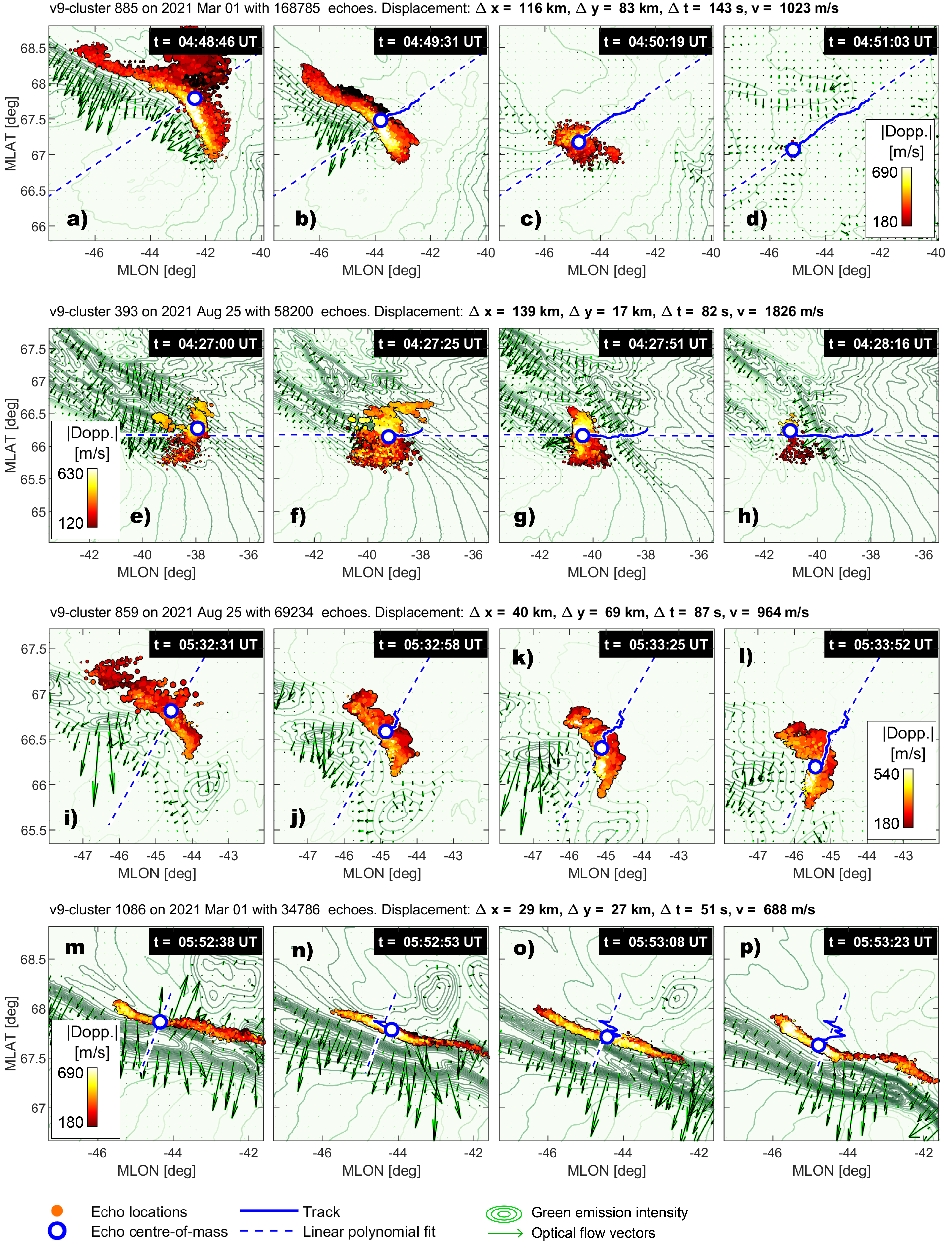}
\caption{\label{fig:cluster1}Detailed spatio-temporal evolution of four \textsc{icebear} echo clusters (red-yellow-white point-clouds, colorscale indicated), as well as their spatial tracks (solid blue lines). The point-clouds are superposed on optical auroral imaging, represented by green contour lines that trace pixel intensity. Green arrows show the optical flow of pixel intensity. A dashed blue line shows linear polynomials fitted to (magnetic) longitudinal and latitudinal components of the tracks, and display the overall direction travelled. The bulk speed inferred from those fits is displayed above each of the three rows.}
\end{figure*}

The final product in this process are the \emph{tracks} traced out by each cluster as it moves through space. In Figure~\ref{fig:tracks} we show two such tracks as solid blue lines. We fit a linear polynomial to the longitudinal and latitudinal positions (dashed blue lines). The coefficient, or slope, in these fits comprise the eastward and poleward bulk velocity respectively, and the 95-percentile (3-sigma) confidence intervals of those fits readily provide errorbars to those bulk velocity estimates. Individual clusters (such as those identified in Figure~\ref{fig:ex}) produce spatial tracks, and the minutiae of  their time histories represent each cluster's spatio-temporal evolution. In the next section we shall present extensive results of such bulk motion determination.

\section{Results} \label{sec:results}

Figure~\ref{fig:cluster1} shows four such time histories, where we plot four equally spaced frames in the life of four selected clusters that were identified on 1 March and 25 August, in 2021, all observed pre-midnight magnetic local time. We display the auroral images (in the green channel) as contour lines of constant emission intensity, with green arrows showing the result of applying the Horn-Schunk optical flow algorithm to the images \cite{barronPerformanceOpticalFlow1994}. The four clusters exhibit bulk motions, and the spatial tracks (solid blue lines) are well approximated by the linear polynomial fits (dashed blue lines), with the notable exception of the bottom row (panels m--p), which we will discuss in the next paragraph. All the clusters of echoes are seen in close association with the spatial distribution of auroral emissions, in accordance with existing observations of auroral turbulence \cite{bahcivanObservationsColocatedOptical2006,huyghebaertPropertiesICEBEARERegion2021,ivarsenTurbulenceAuroras2024}. Likewise, for all the clusters in Figure~\ref{fig:cluster1}, the direction of the straight, dashed blue lines in  points in the overall direction of the aurora. The distribution of specific Doppler shifts, indicated by the color of the echo point-clouds, follow the distribution of auroral emissions, with distinct yellow regions (fast Doppler speeds) appearing close to auroral emissions, where the electric field is expected to maximize \cite{maynardExampleAnticorrelationAuroral1973}. The Doppler speeds of radar echoes in the E-region are expected to correlate with the F-region drifts convection electric field, but are limited in magnitude by instability physics \cite{nielsen_first_1983,fosterSimultaneousObservationsEregion2000,koustovRelationshipVelocityEregion2005}, a topic that we will discuss in the next section.

The first and last row of Figure~\ref{fig:cluster1} corresponds to the two tracks on display in Figure~\ref{fig:tracks}, showing how echo clusters can drift in a clear and coherent way (Figure~\ref{fig:cluster1}a--d) or drifting hither and thither (Figure~\ref{fig:cluster1}m--p), while still tracing out a spatial track that clearly follows the motion of an auroral form. In terms of ionospheric electrodynamics surrounding auroral arcs, we offer the following explanation. In the case of the former, Figure~\ref{fig:cluster1}a--d), the echo cluster and the auroral form both follow an ambient ionospheric drift, which at that time and location (evening auroral oval) pointed in the south-western direction. In the case of the latter, Figure~\ref{fig:cluster1}m--p), we likewise observe an auroral arc that is embedded into an ambient, convection electric field, but this time the auroral arc is unusually intense, producing a local electric field that is strong and variable, effectively \emph{overriding} the ambient field (see, e.g., Figure~7 in \cite{marklund_electric_2009}, Figure~5 in \cite{hosokawaLargeFlowShears2013}, Figure~7 in \cite{gallardo-lacourt_ionospheric_2014}, and Figure~3 in \cite{dubyagin_evidence_2003}). This explanation makes sense if the \textsc{icebear} echo clusters, which are excited by a strong perpendicular electric field \cite{fejerIonosphericIrregularities1980}, follow the motion of electric field source regions.

\section{Discussion}

\begin{figure}
    \centering
    \includegraphics[width=.495\textwidth]{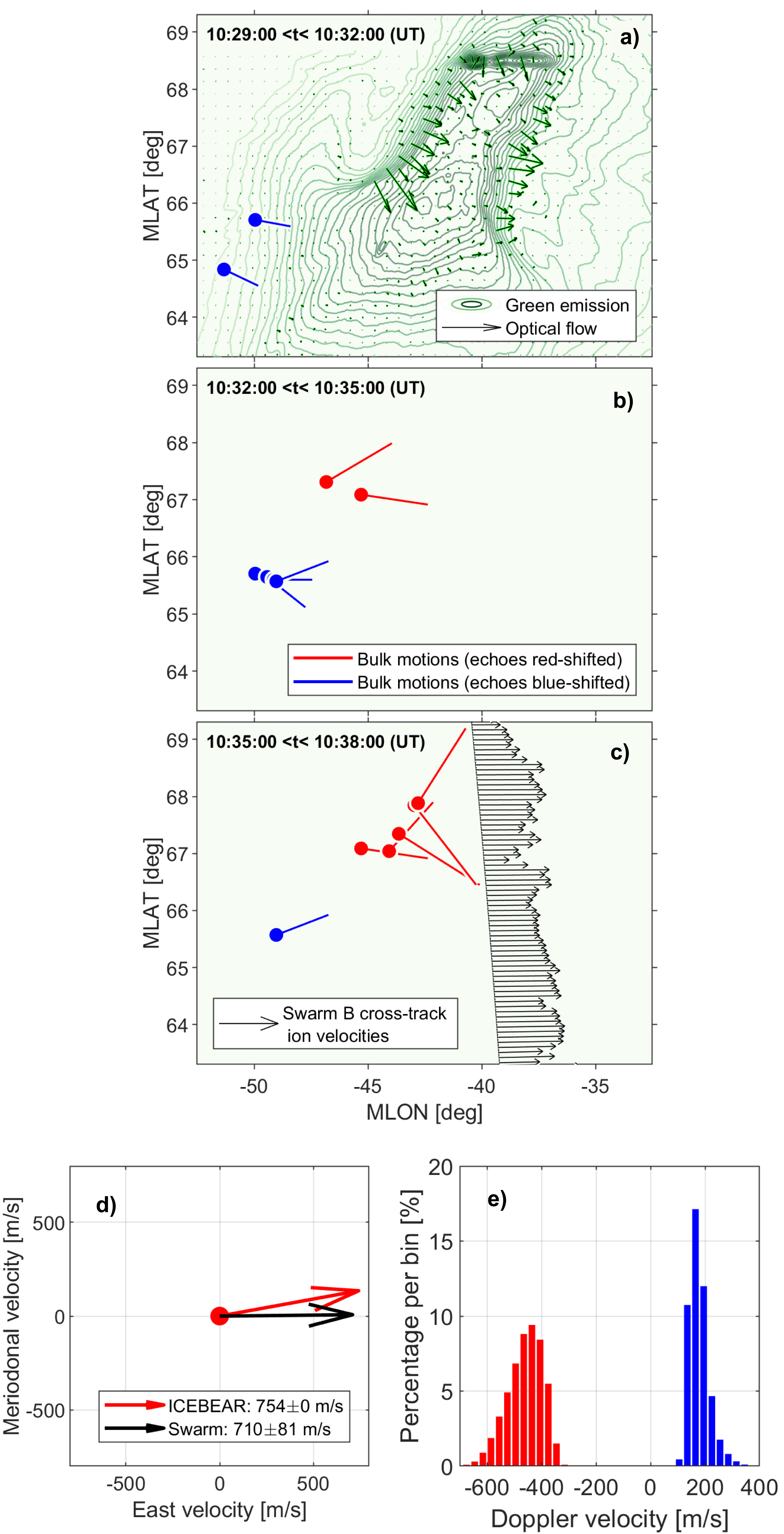}
    \caption{\textbf{Panels a--c):} An early-morning conjunction between ground-based optical imaging of the aurora (green contour lines, with green arrows denoting auroral optical flow), \textsc{icebear} echo cluster tracks (red or blue lines, depending on the Doppler shift of the underlying echoes), and Swarm~B cross-track ion drift observations from the F-region (black arrows). Each of the three panels display consecutive three-minute intervals from 10:29~UT to 10:38~UT on 5 September 2022, of concurrent observations plotted in an MLON-MLAT plane that take field lines curvature between the E- and F-regions into account. \textbf{Panel d)} shows the echo-weighted average \textsc{icebear} bulk motion (red arrow) as well as the average cross-track Swarm~B ion drifts (blue) for the time interval 10:35~UT -- 10:38~UT. \textbf{Panel e)} shows the distributions in Doppler velocity for all the echoes observed during the 9-minute interval, color-coded.}
    \label{fig:swarm}
\end{figure}

The motion of the aurora follows plasma convection and large-scale instability processes in the magnetosphere \cite{voronkov_coupling_1997,uritsky_longitudinally_2009,chastonMotionAurorae2010,ogasawara_azimuthal_2011,miyashitaMagneticFieldEnergetic2021}. The same fields that are involved in those large-scale processes also power the (smaller-scale) wave-particle interactions which ultimately cause particles to precipitate into Earth's atmosphere via pitch-angle scattering \cite{thorne_scattering_2010,kasaharaPulsatingAuroraElectron2018,hosokawaMultipleTimescaleBeats2020} or through parallel electrostatic fields \cite{echimMagnetosphericGeneratorDriving2009,borovskyQuiescentDiscreteAuroral2019,imajo_active_2021}. Since the local fields are, in turn, responsible for producing auroral plasma turbulence \cite{fejerIonosphericIrregularities1980,ivarsenTurbulenceAuroras2024}, it is not surprising that the latter should ultimately follow the motion of magnetospheric source regions, and hence also the general ionospheric convection. In support of the notion that the \textsc{icebear} bulk motions follow the ionospheric plasma convection, we present in Figure~\ref{fig:swarm} a space-ground conjunction with one of the European Space Agency's Swarm satellites. The Swarm mission consists of three low-Earth-orbit (450~km -- 500~km altitude) with a high inclination (87$^\circ$), each equipped with an electric field instrument \cite{friis-christensenSwarmConstellationStudy2006,knudsenThermalIonImagers2017}. The latter computes ion drift velocities from an advanced imaging procedure.

Figure~\ref{fig:swarm}a--c) show three consecutive three-minute intervals during geomagnetic dawn (around 03h magnetic local time) on 5 September 2022. The auroral images (green contour lines) show a large, roughly north-south oriented diffuse auroral form, itself consisting of numerous flashing or pulsating auroral patches that drift steadily in the eastward direction. The pulsating aurora has been shown statistically to follow the large-scale convection pattern, to the extent that it has been proposed as its proxy measurement \cite{yangStatisticalStudyMotion2017}. At first sunlight, on 10:31:15~UT, the \textsc{tre}x \textsc{rgb} camera system was shut off to preserve the light-sensitive sensors. Naturally, \textsc{icebear}, being a radar, continued unaffected to observe the bulk motion of extant echo clusters, of which the majority were pointing in the flow-direction of the diffuse auroral form. At around 10:36:30~UT Swarm~B orbited through \textsc{icebear}'s field-of-view, with the cross-track \emph{in-situ} ion velocity measurements from the F-region at an altitude 500~km exhibiting a steady eastward drift. A video of the event is provided in the Supporting Information, which clearly shows the inferences written here.

Figure~\ref{fig:swarm}d) compares \textsc{icebear}'s echo-weighted average bulk motion with Swarm B's average cross-track ion drift for the interval 10:35~UT -- 10:38~UT, showing excellent agreement in both direction and magnitude. This observation agrees well with the inferences made from Figure~\ref{fig:cluster1}a--l), namely that the \textsc{icebear} echo clusters' general motion and spatial distribution track the aurora. 


The Doppler velocity of the echoes observed during the 5 September 2022 conjunction (distributions of which are shown in Figure~\ref{fig:swarm}e) are organized in two distinct populations. One equatorward moves towards the receiver (blue-shifted) while one poleward moves away from the receiver (red-shifted), possibly indicating a flow reversal . The former is slow ($\sim200$~m/s) while the latter is likely at the ion acoustic velocity ($\sim500$~m/s), both considerably slower than the bulk speeds as well as the F-region ion drifts.

The Doppler speed of the FB waves usually cannot exceed a threshold speed that is achieved by a combination of evolving aspect angles and perpendicular mode-coupling \cite{drexlerNewInsightsNonlocal2002,oppenheim_large-scale_2008,oppenheim_kinetic_2013}. As a result, E-region backscatter echoes from FB waves exhibit their largest  Doppler shifts in the ${\bf E}\times {\bf B}$ direction \cite{koustovRelationshipVelocityEregion2005}, but limited to notional ion acoustic velocity. Simultaneous coherent (E-region) and incoherent (F-region) radar measurements by,  among others, \cite{nielsen_first_1983} and \cite{fosterSimultaneousObservationsEregion2000} demonstrate this important fact. The source regions responsible for exciting FB waves (electric field enhancements) are naturally not limited by these considerations, and, as we explained above, are intimately connected to auroral forms. The explanation for the discrepancy between Doppler and bulk motions then becomes clear: if the source region is moving rapidly, FB waves (each of which move slower than the ion acoustic speed) will be excited along the path of the source region. Each wave is very short-lived (a second or less \cite{ivarsen_deriving_2024}) and so the tracked cluster of echoes will appear to move with the source region.

Curiously, however, Figure~\ref{fig:swarm} also exhibits a few instances where there is also a discrepancy between the Doppler \emph{direction} and the direction of the bulk motion. An opposite direction between radar echoes' bulk motion and Doppler velocities have been observed before \cite{chauUnusualRegionFieldaligned2016}, though without a clear explanation. In our case, the line-of-sight (roughly along the north-south meridian) is likely to be largely perpendicular to the $\boldsymbol{E}\times\boldsymbol{B}$-drift, meaning that there may not be any instability in the line-of-sight direction. FB waves with phase speeds slower than the ion acoustic speed may be indicative of irregularity growth along a direction different from the most unstable direction \cite{oppenheim_large-scale_2008}, and the blue-shifted echoes in Figure~\ref{fig:swarm}e are conspicuously slow, at around 200~m/s. At any rate, a thorough explanation for the directional discrepancy between the Doppler and bulk motions is outside of the scope of the present report, and we defer the matter to a future investigation.

Lastly, we note that Figure~\ref{fig:swarm} contains a total of four distinct measures of velocity. The auroral flow vectors are related to the motion of pulsating, diffuse auroras at dawn, which is known to track the F-region convection \cite{yangStatisticalStudyMotion2017}, though we do not subject the optical flow to any further analysis, as its magnitude depends explicitly on an arbitrary measure of pixel intensity. The Doppler velocities are limited by instability physics \cite{nielsen_first_1983}, dependent on electron temperatures and altitude \cite{st-mauriceRevisitingBehaviorERegion2021,st-mauriceNarrowWidthFarleyBuneman2023}, and are in this instance largely perpendicular to the $\boldsymbol{E}\times\boldsymbol{B}$-drifts. The F-region ion drifts and echo the bulk motions, however, agree in overall direction and magnitude, and are likely both discerning the ionospheric electrodynamics.

\subsection{Electric field enhancements around auroral arcs}

Next, we shall discuss the prominent and important exceptions to the stated rule that the motion of echo clusters follow the motion of auroras. The bottom row of Figure~\ref{fig:cluster1} presents a curious case where \textsc{icebear}'s echo cluster track ostensibly vary considerably in directions perpendicular to the general motion of the aurora. As we suggested in the previous section, if the local field enhancements poleward of the discrete arc in \ref{fig:cluster1}m--p) are strong enough, they will cancel out the ambient drift and cause the local $\boldsymbol{E}\times\boldsymbol{B}$-direction to fluctuate. Despite this, the average direction over 51~seconds (the dashed blue lines in Figure~\ref{fig:cluster1}m--p) is perfectly perpendicular to the discrete arc and pointing in the direction of the auroral flow vectors. In other words, in this case, the average bulk motion still uncovers a general drift direction consistent with the expectations for the large-scale electric field as observed in the pre-midnight auroral region. 

\begin{figure}
    \centering
    \includegraphics[width=.495\textwidth]{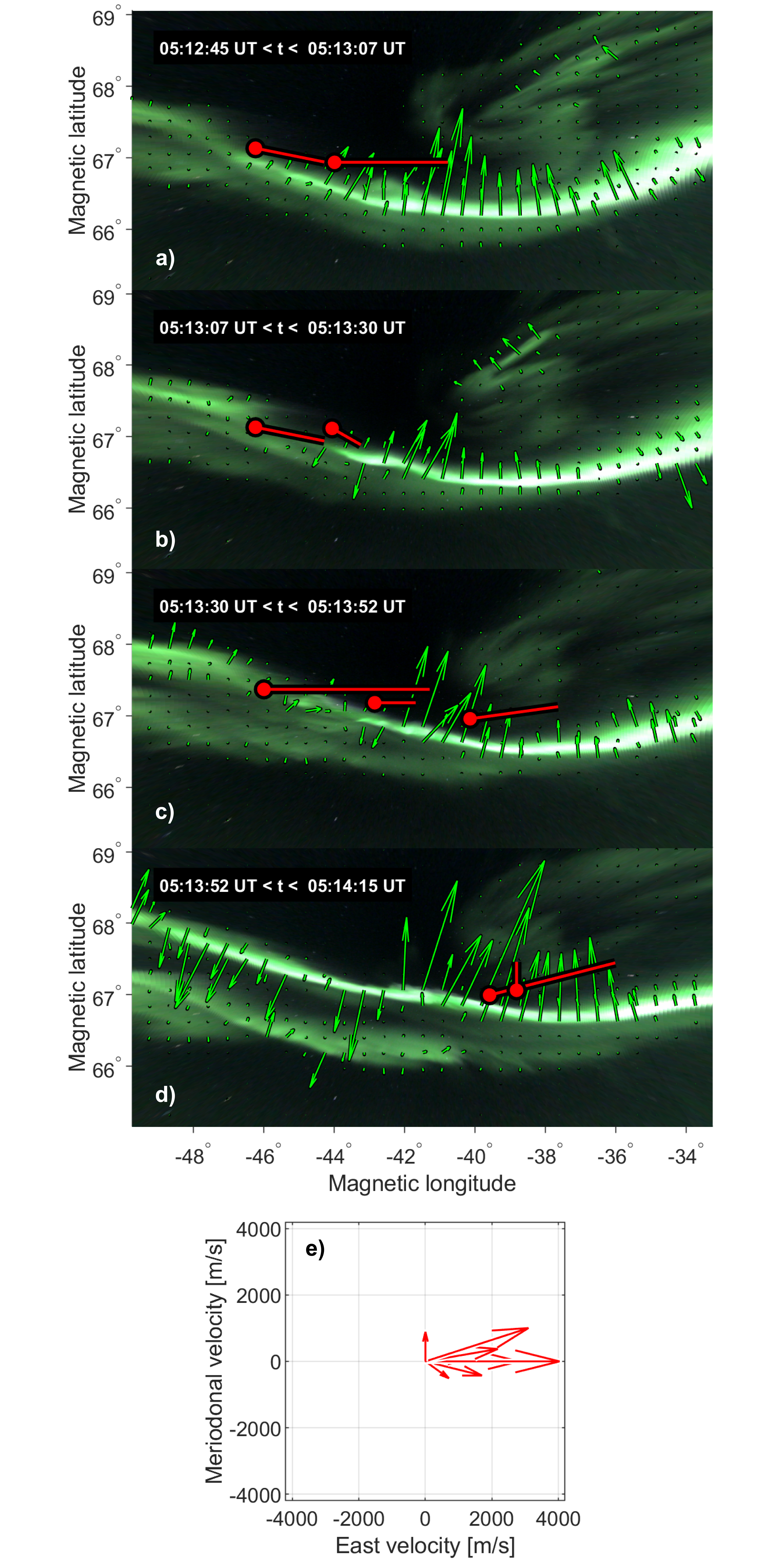}
    \caption{\textbf{Panels a--c):} \textsc{icebear} echo tracks (solid red lines) and their linear polynomial fits (dashed red lines), superposed on auroral images, with green arrows denoting auroral optical flow. Each panel encompasses around 20~seconds of data. Panel d) shows the various \textsc{icebear} bulk motion directions and magnitudes for the entire 90-second interval.}
    \label{fig:tracks_avg}
\end{figure}

Figure~\ref{fig:tracks_avg} shows an even more extreme example of the foregoing, where the tracks point parallel to an intense, undulating auroral arc, reaching the considerable bulk speed of 4100~m/s. Picking up the thread from the previous section, we note that the local electric field around these unusually intense precipitation regions is strong enough to completely override the ambient drift, causing the bulk motions to be parallel to the arc and  extremely fast. In this case, the local electric field, in principle a linear superposition of that field and the ambient, is almost wholly produced and maintained by  energetic particle precipitation. Examples of such strong perpendicular fields pointing towards auroral arcs have been observed \emph{in-situ} \cite{dubyagin_evidence_2003}, with incoherent scatter radar in conjunction with sounding rockets \cite{opgenoorthRegionsStronglyEnhanced1990}, as well as specifically in the case of intense, undulating auroral arcs \cite{lanchester_relationship_1996}. Figure~\ref{fig:tracks_avg} is entirely consistent with these earlier observations, strengthening the notion that the echo bulk motions seem to track regions of enhanced electric fields.


\subsection{Validation}

\begin{figure}
    \centering
    \includegraphics[width=.39\textwidth]{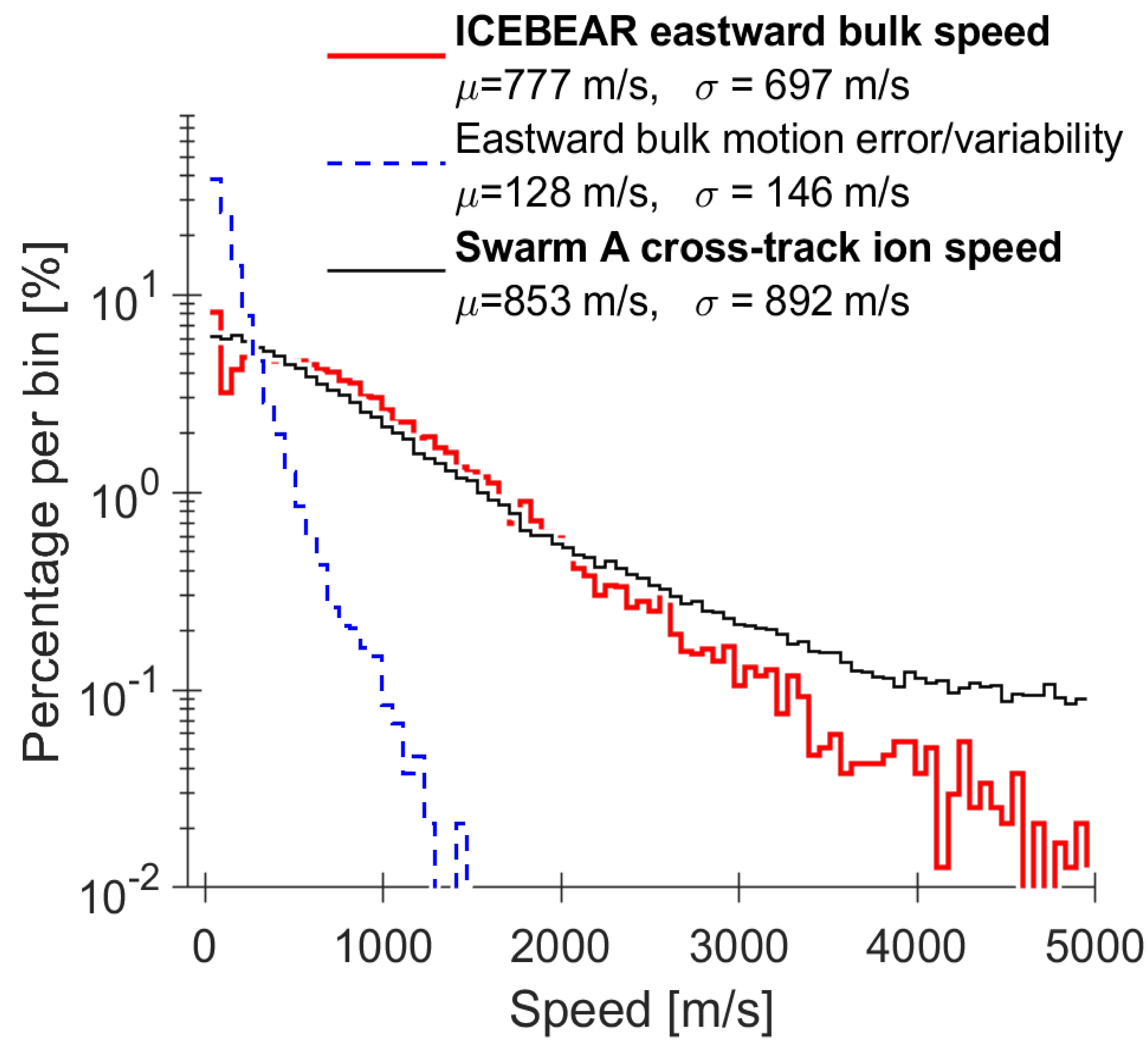}
    \caption{A statistical aggregate of the \textsc{icebear} eastward echo bulk speeds (red and blue) compared to the Swarm~A cross-track ion drifts (black). The data is collected between the magnetic local times 16h and 08h and the magnetic latitudes $62^\circ$ and $70^\circ$, and the Swarm-data is collected during conditions when the auroral electrojet index (the SME-index \cite{gjerloevSuperMAGDataProcessing2012}) exceeds 150~nT, as \textsc{icebear} echoes are not observed when the SME-index is lower than 150~nT. Mean ($\mu$) and standard deviation ($\sigma$) values are indicated. }
    \label{fig:stats}
\end{figure}

Our proposed method of tracking turbulent structures is still in its infancy and has yet to be perfected. The resulting tracks and the bulk velocities they imply must, at present, be carefully assessed before conclusions are drawn about their physical significance. Nevertheless, we observe that the bulk motions appear to follow the motion of the echo source regions quite closely. Without implying at present that the two are equivalent, we can compare the echo bulk motion to the $\boldsymbol{E}\times\boldsymbol{B}$-motion in the ionosphere. To that end, and to present a preliminary statistical overview of the bulk motions, we show with a solid red line in Figure~\ref{fig:stats} a distribution of all eastward bulk motions observed during January 2020 -- May 2023, observed at night between the magnetic local times of 16h and 08h, where the only selection criterion is that the variability (see Figure~\ref{fig:tracks}) not exceed a value one third of the bulk speed. We compared this to the Swarm~A cross-track ion drifts observed in the same magnetic local time interval, and between the same magnetic latitudes ($62^\circ$ and $70^\circ$), the distribution of which is shown in Figure~\ref{fig:stats} with a solid black line. In addition, we plot with a dashed blue line the distribution of the bulk motion variability. Mean ($\mu$) and standard deviation ($\sigma$) for all three distributions are indicated.

The general similarity in the shape of the red and black distributions in Figure~\ref{fig:stats}, and the general agreement in mean and standard deviation for those distributions, indicate that the various bulk motions we can observe in a very large dataset (some 22,000 tracked echo clusters), are reasonably constrained by the physics of ionospheric electrodynamics. The error, or variability, in the inferred bulk speeds (dashed blue line in Figure~\ref{fig:stats}) are much smaller in magnitude than the bulk speeds themselves, although they are considerably larger than error margins expected of precision measurements of the auroral motions \cite{haerendel_proper_1993}. In a related publication, \cite{ivarsen_deriving_2024}, the notion that the echo bulk motions may in fact track electric field source regions is more thoroughly motivated, and better demonstrated, and we suffice in the present report to note the statistical similarity between the bulk motions and the F-region ion drifts. 

Some caveats should, however, be noted. Critical tracking situations may occur when the clustering algorithm decides to drop or include echo regions, effectively causing bifurcation or merging events in the dataset. During such occurrences the median echo location may suddenly shift, causing a jump in the track. This jump will go on to form the track's subsequent trajectory after merging or bifurcation, and as such is not necessarily unphysical, but it places upon the algorithm what may appear as an arbitrary means of deciding when clusters merge or bifurcate. A point of potential improvement for future development lies in the possibility of clustering in higher dimensions \cite{sander_density-based_1998}. Here, time could be included as a dimension, removing the need for timestep-by-timestep matching of clusters to their preceding form. Doppler speed and signal-to-noise ratio, or even altitude, may also be added to the dimensions to cluster in. This can be done, for example, through dimensionality reduction by principal component analysis.

The present paper has applied density-based clustering, an unsupervised machine learning method for data mining \cite{sander_density-based_2010}, part of a growing field of analysis that have garnered interest also from the space weather community \cite{camporeale_machine_2018}. Other clustering methods such as the $k$-nearest neighbour classification has recently been applied to forecast global total electron content (TEC) measurements \cite{monte-moreno_forecast_2022}, and related neural network-based methods have been applied to forecast and predict geomagnetic storms for decades \cite{valdivia_prediction_1996}, as well as in recent implementations of image classification using data from the Swarm mission \cite{antonopoulou_convolutional_2022}. In the future, supervised learning strategies may aide our method in determining optical parameters for echo clustering,

\section{Concluding remarks}


\begin{figure*}
\includegraphics[width=\textwidth]{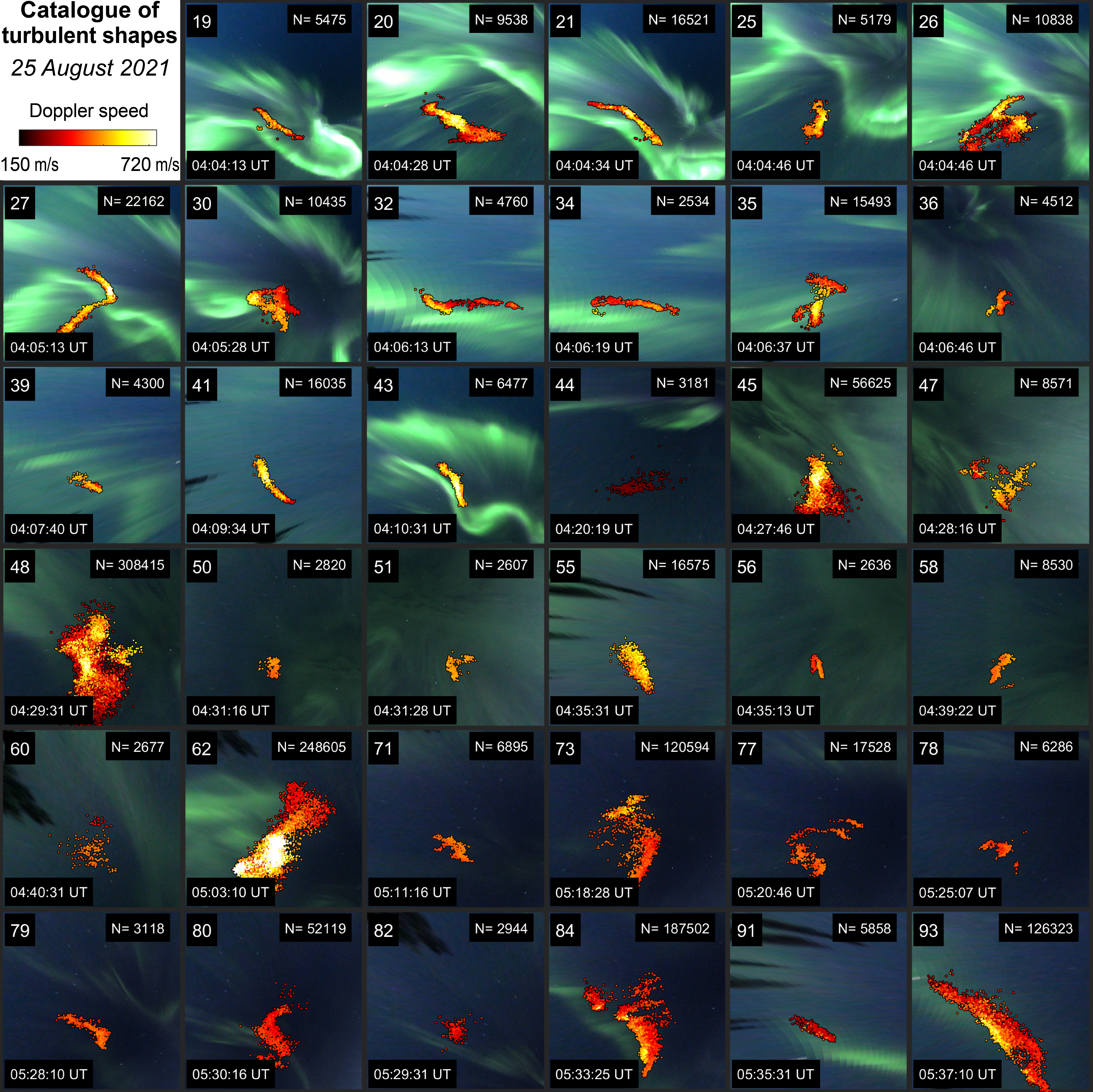}
\caption{\label{fig:menagerie}35 clusters of \textsc{icebear}~3D echoes identified on 25 August 2021. Each panel covers an area of $266$~km~$\times~266$~km and shows the cluster as it appeared at its most echo-numerous timestep.  The clusters are superposed on the \textsc{tre}x \textsc{rgb} auroral images observed at that time. A red-yellow-white color scale indicates the absolute Doppler speed of the echoes.}
\end{figure*}

We have applied industry-proven point-cloud analysis techniques to automatically mine large datasets of interferometry radar data. The method entails point-cloud clustering in two dimensions (in our case, the local plane perpendicular to Earth's magnetic field lines). We track clusters of radar echoes through time by a simple similarity test from timestep-to-timestep. The resulting time histories of echo clusters trace out distinct spatial tracks, from which we derive a bulk motion. Through several case studies,  we show that this motion tends to follow the motion of auroral forms. Notable exceptions to this rule occur when especially intense auroral arcs cause a fast, peculiar motion parallel to those arcs, at which point our method of extracting bulk motions from identified clusters of echoes replicates results from previous studies of auroral electric field enhancements \cite{opgenoorthRegionsStronglyEnhanced1990,lanchester_relationship_1996,dubyagin_evidence_2003}.

Our method, which is fully automatic, can be used to mine point-cloud data for irregular and dynamic clustering and flag this data for subsequent analyses. To illustrate the usefulness of our algorithm's automatic nature, we have produced in Figure~\ref{fig:menagerie} a serial collection of 35 \textsc{icebear} echo clusters that were observed between 04:00~UT and 05:40~UT on 28 August 2021, each superposed on optical images of the aurora. The echo clusters are represented by snapshots taken at a prominent point during their evolution. For each of the 35 shapes, then, there readily exists a detailed frame-by-frame evolution akin to those presented in Figures~\ref{fig:cluster1}.

With \textsc{icebear}'s new observations of the radar aurora bulk motions come new opportunities to study the magnetosphere-ionosphere coupling that lies at the heart of such motions, and to shed further light on the understudied E-region ionosphere.
\vspace{8pt}

\section*{Acknowledgements}
This work is supported in part by Research Council of Norway (RCN) grant 324859. We acknowledge the support of the Canadian Space Agency (CSA) [20SUGOICEB], the Canada Foundation for Innovation (CFI) John R. Evans Leaders Fund [32117], the Natural Science and Engineering Research Council (NSERC), the International Space Mission Training Program supported by the Collaborative Research and Training Experience (CREATE) [479771-2016], the Discovery grants program [RGPIN-2019-19135]; and the Digital Research Alliance of Canada [RRG-FT2109]. DRH was funded during this study through a UiT The Arctic University of Norway contribution to the EISCAT\_3D project funded by RCN grant 245683. MFI is grateful to M. Oppenheim, Y. Dimant, F. Lind, P. Erickson, and Y. Miyashita for stimilating discussions.

\section*{Data Availability}
\textsc{icebear} 3D echo data for 2020, 2021 is published with DOI \texttt{10.5281/zenodo.7509022}. \textsc{tre}x optical data can be accessed at \url{https://data.phys.ucalgary.ca/}. Data from the European Space Agency's Swarm mission can accessed at \texttt{https://swarm-diss.eo.esa.int/}.


\begin{thebibliography}{70}%
\makeatletter
\providecommand \@ifxundefined [1]{%
 \@ifx{#1\undefined}
}%
\providecommand \@ifnum [1]{%
 \ifnum #1\expandafter \@firstoftwo
 \else \expandafter \@secondoftwo
 \fi
}%
\providecommand \@ifx [1]{%
 \ifx #1\expandafter \@firstoftwo
 \else \expandafter \@secondoftwo
 \fi
}%
\providecommand \natexlab [1]{#1}%
\providecommand \enquote  [1]{``#1''}%
\providecommand \bibnamefont  [1]{#1}%
\providecommand \bibfnamefont [1]{#1}%
\providecommand \citenamefont [1]{#1}%
\providecommand \href@noop [0]{\@secondoftwo}%
\providecommand \href [0]{\begingroup \@sanitize@url \@href}%
\providecommand \@href[1]{\@@startlink{#1}\@@href}%
\providecommand \@@href[1]{\endgroup#1\@@endlink}%
\providecommand \@sanitize@url [0]{\catcode `\\12\catcode `\$12\catcode `\&12\catcode `\#12\catcode `\^12\catcode `\_12\catcode `\%12\relax}%
\providecommand \@@startlink[1]{}%
\providecommand \@@endlink[0]{}%
\providecommand \url  [0]{\begingroup\@sanitize@url \@url }%
\providecommand \@url [1]{\endgroup\@href {#1}{\urlprefix }}%
\providecommand \urlprefix  [0]{URL }%
\providecommand \Eprint [0]{\href }%
\providecommand \doibase [0]{https://doi.org/}%
\providecommand \selectlanguage [0]{\@gobble}%
\providecommand \bibinfo  [0]{\@secondoftwo}%
\providecommand \bibfield  [0]{\@secondoftwo}%
\providecommand \translation [1]{[#1]}%
\providecommand \BibitemOpen [0]{}%
\providecommand \bibitemStop [0]{}%
\providecommand \bibitemNoStop [0]{.\EOS\space}%
\providecommand \EOS [0]{\spacefactor3000\relax}%
\providecommand \BibitemShut  [1]{\csname bibitem#1\endcsname}%
\let\auto@bib@innerbib\@empty
\bibitem [{\citenamefont {Birant}\ and\ \citenamefont {Kut}(2007)}]{birantSTDBSCANAlgorithmClustering2007}%
  \BibitemOpen
  \bibfield  {author} {\bibinfo {author} {\bibfnamefont {D.}~\bibnamefont {Birant}}\ and\ \bibinfo {author} {\bibfnamefont {A.}~\bibnamefont {Kut}},\ }\bibfield  {title} {\bibinfo {title} {{{ST-DBSCAN}}: {{An}} algorithm for clustering spatial--temporal data},\ }\href {https://doi.org/10.1016/j.datak.2006.01.013} {\bibfield  {journal} {\bibinfo  {journal} {Data \& Knowledge Engineering}\ }\bibinfo {series} {Intelligent {{Data Mining}}},\ \textbf {\bibinfo {volume} {60}},\ \bibinfo {pages} {208} (\bibinfo {year} {2007})}\BibitemShut {NoStop}%
\bibitem [{\citenamefont {McDaniel}\ \emph {et~al.}(2012)\citenamefont {McDaniel}, \citenamefont {Nishihata}, \citenamefont {Brooks}, \citenamefont {Salesses},\ and\ \citenamefont {Iagnemma}}]{mcdanielTerrainClassificationIdentification2012}%
  \BibitemOpen
  \bibfield  {author} {\bibinfo {author} {\bibfnamefont {M.~W.}\ \bibnamefont {McDaniel}}, \bibinfo {author} {\bibfnamefont {T.}~\bibnamefont {Nishihata}}, \bibinfo {author} {\bibfnamefont {C.~A.}\ \bibnamefont {Brooks}}, \bibinfo {author} {\bibfnamefont {P.}~\bibnamefont {Salesses}},\ and\ \bibinfo {author} {\bibfnamefont {K.}~\bibnamefont {Iagnemma}},\ }\bibfield  {title} {\bibinfo {title} {Terrain classification and identification of tree stems using ground-based {{LiDAR}}},\ }\href {https://doi.org/10.1002/rob.21422} {\bibfield  {journal} {\bibinfo  {journal} {Journal of Field Robotics}\ }\textbf {\bibinfo {volume} {29}},\ \bibinfo {pages} {891} (\bibinfo {year} {2012})}\BibitemShut {NoStop}%
\bibitem [{\citenamefont {Li}\ \emph {et~al.}(2017)\citenamefont {Li}, \citenamefont {Bao}, \citenamefont {Han}, \citenamefont {Pan}, \citenamefont {Pan}, \citenamefont {Zhang},\ and\ \citenamefont {Wang}}]{liRealtimeSelfdrivingCar2017}%
  \BibitemOpen
  \bibfield  {author} {\bibinfo {author} {\bibfnamefont {J.}~\bibnamefont {Li}}, \bibinfo {author} {\bibfnamefont {H.}~\bibnamefont {Bao}}, \bibinfo {author} {\bibfnamefont {X.}~\bibnamefont {Han}}, \bibinfo {author} {\bibfnamefont {F.}~\bibnamefont {Pan}}, \bibinfo {author} {\bibfnamefont {W.}~\bibnamefont {Pan}}, \bibinfo {author} {\bibfnamefont {F.}~\bibnamefont {Zhang}},\ and\ \bibinfo {author} {\bibfnamefont {D.}~\bibnamefont {Wang}},\ }\bibfield  {title} {\bibinfo {title} {Real-time self-driving car navigation and obstacle avoidance using mobile {{3D}} laser scanner and {{GNSS}}},\ }\href {https://doi.org/10.1007/s11042-016-4211-7} {\bibfield  {journal} {\bibinfo  {journal} {Multimedia Tools and Applications}\ }\textbf {\bibinfo {volume} {76}},\ \bibinfo {pages} {23017} (\bibinfo {year} {2017})}\BibitemShut {NoStop}%
\bibitem [{\citenamefont {Hodgson}\ \emph {et~al.}(2005)\citenamefont {Hodgson}, \citenamefont {Jensen}, \citenamefont {Raber}, \citenamefont {Tullis}, \citenamefont {Davis}, \citenamefont {Thompson},\ and\ \citenamefont {Schuckman}}]{hodgsonEvaluationLidarderivedElevation2005}%
  \BibitemOpen
  \bibfield  {author} {\bibinfo {author} {\bibfnamefont {M.~E.}\ \bibnamefont {Hodgson}}, \bibinfo {author} {\bibfnamefont {J.}~\bibnamefont {Jensen}}, \bibinfo {author} {\bibfnamefont {G.}~\bibnamefont {Raber}}, \bibinfo {author} {\bibfnamefont {J.}~\bibnamefont {Tullis}}, \bibinfo {author} {\bibfnamefont {B.~A.}\ \bibnamefont {Davis}}, \bibinfo {author} {\bibfnamefont {G.}~\bibnamefont {Thompson}},\ and\ \bibinfo {author} {\bibfnamefont {K.}~\bibnamefont {Schuckman}},\ }\bibfield  {title} {\bibinfo {title} {An {{Evaluation}} of {{Lidar-derived Elevation}} and {{Terrain Slope}} in {{Leaf-off Conditions}}},\ }\href {https://doi.org/10.14358/PERS.71.7.817} {\bibfield  {journal} {\bibinfo  {journal} {Photogrammetric Engineering \& Remote Sensing}\ }\textbf {\bibinfo {volume} {71}},\ \bibinfo {pages} {817} (\bibinfo {year} {2005})}\BibitemShut {NoStop}%
\bibitem [{\citenamefont {Schindling}\ and\ \citenamefont {Gibbes}(2014)}]{schindlingLiDARToolArchaeological2014}%
  \BibitemOpen
  \bibfield  {author} {\bibinfo {author} {\bibfnamefont {J.}~\bibnamefont {Schindling}}\ and\ \bibinfo {author} {\bibfnamefont {C.}~\bibnamefont {Gibbes}},\ }\bibfield  {title} {\bibinfo {title} {{{LiDAR}} as a tool for archaeological research: A case study},\ }\href {https://doi.org/10.1007/s12520-014-0178-3} {\bibfield  {journal} {\bibinfo  {journal} {Archaeological and Anthropological Sciences}\ }\textbf {\bibinfo {volume} {6}},\ \bibinfo {pages} {411} (\bibinfo {year} {2014})}\BibitemShut {NoStop}%
\bibitem [{\citenamefont {Lozinsky}\ \emph {et~al.}(2022)\citenamefont {Lozinsky}, \citenamefont {Hussey}, \citenamefont {McWilliams}, \citenamefont {Huyghebaert},\ and\ \citenamefont {Galeschuk}}]{lozinskyICEBEAR3DLowElevation2022}%
  \BibitemOpen
  \bibfield  {author} {\bibinfo {author} {\bibfnamefont {A.}~\bibnamefont {Lozinsky}}, \bibinfo {author} {\bibfnamefont {G.}~\bibnamefont {Hussey}}, \bibinfo {author} {\bibfnamefont {K.}~\bibnamefont {McWilliams}}, \bibinfo {author} {\bibfnamefont {D.}~\bibnamefont {Huyghebaert}},\ and\ \bibinfo {author} {\bibfnamefont {D.}~\bibnamefont {Galeschuk}},\ }\bibfield  {title} {\bibinfo {title} {{{ICEBEAR-3D}}: {{A Low Elevation Imaging Radar Using}} a {{Non-Uniform Coplanar Receiver Array}} for {{E Region Observations}}},\ }\href {https://doi.org/10.1029/2021RS007358} {\bibfield  {journal} {\bibinfo  {journal} {Radio Science}\ }\textbf {\bibinfo {volume} {57}},\ \bibinfo {pages} {e2021RS007358} (\bibinfo {year} {2022})}\BibitemShut {NoStop}%
\bibitem [{\citenamefont {Ivarsen}\ \emph {et~al.}(2023{\natexlab{a}})\citenamefont {Ivarsen}, \citenamefont {{St-Maurice}}, \citenamefont {Hussey}, \citenamefont {Galeschuk}, \citenamefont {Lozinsky}, \citenamefont {Pitzel},\ and\ \citenamefont {McWilliams}}]{ivarsenAlgorithmSeparateIonospheric2023}%
  \BibitemOpen
  \bibfield  {author} {\bibinfo {author} {\bibfnamefont {M.~F.}\ \bibnamefont {Ivarsen}}, \bibinfo {author} {\bibfnamefont {J.-P.}\ \bibnamefont {{St-Maurice}}}, \bibinfo {author} {\bibfnamefont {G.~C.}\ \bibnamefont {Hussey}}, \bibinfo {author} {\bibfnamefont {D.}~\bibnamefont {Galeschuk}}, \bibinfo {author} {\bibfnamefont {A.}~\bibnamefont {Lozinsky}}, \bibinfo {author} {\bibfnamefont {B.}~\bibnamefont {Pitzel}},\ and\ \bibinfo {author} {\bibfnamefont {K.~A.}\ \bibnamefont {McWilliams}},\ }\bibfield  {title} {\bibinfo {title} {An {{Algorithm}} to {{Separate Ionospheric Turbulence Radar Echoes From Those}} of {{Meteor Trails}} in {{Large Data Sets}}},\ }\href {https://doi.org/10.1029/2022JA031050} {\bibfield  {journal} {\bibinfo  {journal} {Journal of Geophysical Research: Space Physics}\ }\textbf {\bibinfo {volume} {128}},\ \bibinfo {pages} {e2022JA031050} (\bibinfo {year} {2023}{\natexlab{a}})}\BibitemShut {NoStop}%
\bibitem [{\citenamefont {Ivarsen}\ \emph {et~al.}(2023{\natexlab{b}})\citenamefont {Ivarsen}, \citenamefont {Lozinsky}, \citenamefont {{St-Maurice}}, \citenamefont {Spicher}, \citenamefont {Huyghebaert}, \citenamefont {Hussey}, \citenamefont {Galeschuk}, \citenamefont {Pitzel},\ and\ \citenamefont {Vierinen}}]{ivarsenDistributionSmallScaleIrregularities2023}%
  \BibitemOpen
  \bibfield  {author} {\bibinfo {author} {\bibfnamefont {M.~F.}\ \bibnamefont {Ivarsen}}, \bibinfo {author} {\bibfnamefont {A.}~\bibnamefont {Lozinsky}}, \bibinfo {author} {\bibfnamefont {J.-P.}\ \bibnamefont {{St-Maurice}}}, \bibinfo {author} {\bibfnamefont {A.}~\bibnamefont {Spicher}}, \bibinfo {author} {\bibfnamefont {D.}~\bibnamefont {Huyghebaert}}, \bibinfo {author} {\bibfnamefont {G.~C.}\ \bibnamefont {Hussey}}, \bibinfo {author} {\bibfnamefont {D.}~\bibnamefont {Galeschuk}}, \bibinfo {author} {\bibfnamefont {B.}~\bibnamefont {Pitzel}},\ and\ \bibinfo {author} {\bibfnamefont {J.}~\bibnamefont {Vierinen}},\ }\bibfield  {title} {\bibinfo {title} {The {{Distribution}} of {{Small-Scale Irregularities}} in the {{E-Region}}, and {{Its Tendency}} to {{Match}} the {{Spectrum}} of {{Field-Aligned Current Structures}} in the {{F-Region}}},\ }\href {https://doi.org/10.1029/2022JA031233} {\bibfield  {journal} {\bibinfo  {journal} {Journal of Geophysical Research: Space Physics}\ }\textbf {\bibinfo {volume} {128}},\
  \bibinfo {pages} {e2022JA031233} (\bibinfo {year} {2023}{\natexlab{b}})}\BibitemShut {NoStop}%
\bibitem [{\citenamefont {Ivarsen}\ \emph {et~al.}(2024{\natexlab{a}})\citenamefont {Ivarsen}, \citenamefont {Gilles}, \citenamefont {Huyghebaert}, \citenamefont {{St-Maurice}}, \citenamefont {Lozinsky}, \citenamefont {Galeschuk},\ and\ \citenamefont {Hussey}}]{ivarsen_turbulence_embedded_2024}%
  \BibitemOpen
  \bibfield  {author} {\bibinfo {author} {\bibfnamefont {M.~F.}\ \bibnamefont {Ivarsen}}, \bibinfo {author} {\bibfnamefont {D.~M.}\ \bibnamefont {Gilles}}, \bibinfo {author} {\bibfnamefont {D.~R.}\ \bibnamefont {Huyghebaert}}, \bibinfo {author} {\bibfnamefont {J.~P.}\ \bibnamefont {{St-Maurice}}}, \bibinfo {author} {\bibfnamefont {A.}~\bibnamefont {Lozinsky}}, \bibinfo {author} {\bibfnamefont {D.}~\bibnamefont {Galeschuk}},\ and\ \bibinfo {author} {\bibfnamefont {G.}~\bibnamefont {Hussey}},\ }\bibfield  {title} {\bibinfo {title} {Turbulence {{Embedded}} into the {{Ionosphere}} by {{Electromagnetic Waves}}},\ }\href@noop {} {\bibfield  {journal} {\bibinfo  {journal} {Accepted by JGR: Space Physics}\ } (\bibinfo {year} {2024}{\natexlab{a}})}\BibitemShut {NoStop}%
\bibitem [{\citenamefont {Fejer}\ and\ \citenamefont {Providakes}(1987)}]{fejerHighLatitudeEregion1987}%
  \BibitemOpen
  \bibfield  {author} {\bibinfo {author} {\bibfnamefont {B.}~\bibnamefont {Fejer}}\ and\ \bibinfo {author} {\bibfnamefont {J.}~\bibnamefont {Providakes}},\ }\bibfield  {title} {\bibinfo {title} {High latitude {{E-region}} irregularities: {{New}} results},\ }\href {https://doi.org/10.1088/0031-8949/1987/T18/018} {\bibfield  {journal} {\bibinfo  {journal} {Physica Scripta}\ }\textbf {\bibinfo {volume} {T18}},\ \bibinfo {pages} {167} (\bibinfo {year} {1987})}\BibitemShut {NoStop}%
\bibitem [{\citenamefont {Ochi}(1998)}]{ochiOceanWavesStochastic1998}%
  \BibitemOpen
  \bibfield  {author} {\bibinfo {author} {\bibfnamefont {M.~K.}\ \bibnamefont {Ochi}},\ }\href {https://doi.org/10.1017/CBO9780511529559} {\emph {\bibinfo {title} {Ocean {{Waves}}: {{The Stochastic Approach}}}}},\ Cambridge {{Ocean Technology Series}}\ (\bibinfo  {publisher} {Cambridge University Press},\ \bibinfo {address} {Cambridge},\ \bibinfo {year} {1998})\BibitemShut {NoStop}%
\bibitem [{\citenamefont {Huba}\ \emph {et~al.}(1985)\citenamefont {Huba}, \citenamefont {Hassam}, \citenamefont {Schwartz},\ and\ \citenamefont {Keskinen}}]{hubaIonosphericTurbulenceInterchange1985}%
  \BibitemOpen
  \bibfield  {author} {\bibinfo {author} {\bibfnamefont {J.~D.}\ \bibnamefont {Huba}}, \bibinfo {author} {\bibfnamefont {A.~B.}\ \bibnamefont {Hassam}}, \bibinfo {author} {\bibfnamefont {I.~B.}\ \bibnamefont {Schwartz}},\ and\ \bibinfo {author} {\bibfnamefont {M.~J.}\ \bibnamefont {Keskinen}},\ }\bibfield  {title} {\bibinfo {title} {Ionospheric turbulence: {{Interchange}} instabilities and chaotic fluid behavior},\ }\href {https://doi.org/10.1029/GL012i001p00065} {\bibfield  {journal} {\bibinfo  {journal} {Geophysical Research Letters}\ }\textbf {\bibinfo {volume} {12}},\ \bibinfo {pages} {65} (\bibinfo {year} {1985})}\BibitemShut {NoStop}%
\bibitem [{\citenamefont {Dungey}(1961)}]{dungeyInterplanetaryMagneticField1961}%
  \BibitemOpen
  \bibfield  {author} {\bibinfo {author} {\bibfnamefont {J.~W.}\ \bibnamefont {Dungey}},\ }\bibfield  {title} {\bibinfo {title} {Interplanetary {{Magnetic Field}} and the {{Auroral Zones}}},\ }\href {https://doi.org/10.1103/PhysRevLett.6.47} {\bibfield  {journal} {\bibinfo  {journal} {Physical Review Letters}\ }\textbf {\bibinfo {volume} {6}},\ \bibinfo {pages} {47} (\bibinfo {year} {1961})}\BibitemShut {NoStop}%
\bibitem [{\citenamefont {Cowley}(2000)}]{cowleyTUTORIALMagnetosphereIonosphereInteractions2000}%
  \BibitemOpen
  \bibfield  {author} {\bibinfo {author} {\bibfnamefont {S.~W.~H.}\ \bibnamefont {Cowley}},\ }\bibfield  {title} {\bibinfo {title} {{{TUTORIAL}}: {{Magnetosphere-Ionosphere Interactions}}: {{A Tutorial Review}}},\ }\href {https://doi.org/10.1029/GM118p0091} {\bibfield  {journal} {\bibinfo  {journal} {Washington DC American Geophysical Union Geophysical Monograph Series}\ }\textbf {\bibinfo {volume} {118}},\ \bibinfo {pages} {91} (\bibinfo {year} {2000})}\BibitemShut {NoStop}%
\bibitem [{\citenamefont {Keiling}\ \emph {et~al.}(2019)\citenamefont {Keiling}, \citenamefont {Thaller}, \citenamefont {Wygant},\ and\ \citenamefont {Dombeck}}]{keilingAssessingGlobalAlfven2019}%
  \BibitemOpen
  \bibfield  {author} {\bibinfo {author} {\bibfnamefont {A.}~\bibnamefont {Keiling}}, \bibinfo {author} {\bibfnamefont {S.}~\bibnamefont {Thaller}}, \bibinfo {author} {\bibfnamefont {J.}~\bibnamefont {Wygant}},\ and\ \bibinfo {author} {\bibfnamefont {J.}~\bibnamefont {Dombeck}},\ }\bibfield  {title} {\bibinfo {title} {Assessing the global {{Alfv{\'e}n}} wave power flow into and out of the auroral acceleration region during geomagnetic storms},\ }\href {https://doi.org/10.1126/sciadv.aav8411} {\bibfield  {journal} {\bibinfo  {journal} {Science Advances}\ }\textbf {\bibinfo {volume} {5}},\ \bibinfo {pages} {eaav8411} (\bibinfo {year} {2019})}\BibitemShut {NoStop}%
\bibitem [{\citenamefont {Wiltberger}\ \emph {et~al.}(2017)\citenamefont {Wiltberger}, \citenamefont {Merkin}, \citenamefont {Zhang}, \citenamefont {Toffoletto}, \citenamefont {Oppenheim}, \citenamefont {Wang}, \citenamefont {Lyon}, \citenamefont {Liu}, \citenamefont {Dimant}, \citenamefont {Sitnov},\ and\ \citenamefont {Stephens}}]{wiltbergerEffectsElectrojetTurbulence2017}%
  \BibitemOpen
  \bibfield  {author} {\bibinfo {author} {\bibfnamefont {M.}~\bibnamefont {Wiltberger}}, \bibinfo {author} {\bibfnamefont {V.}~\bibnamefont {Merkin}}, \bibinfo {author} {\bibfnamefont {B.}~\bibnamefont {Zhang}}, \bibinfo {author} {\bibfnamefont {F.}~\bibnamefont {Toffoletto}}, \bibinfo {author} {\bibfnamefont {M.}~\bibnamefont {Oppenheim}}, \bibinfo {author} {\bibfnamefont {W.}~\bibnamefont {Wang}}, \bibinfo {author} {\bibfnamefont {J.~G.}\ \bibnamefont {Lyon}}, \bibinfo {author} {\bibfnamefont {J.}~\bibnamefont {Liu}}, \bibinfo {author} {\bibfnamefont {Y.}~\bibnamefont {Dimant}}, \bibinfo {author} {\bibfnamefont {M.~I.}\ \bibnamefont {Sitnov}},\ and\ \bibinfo {author} {\bibfnamefont {G.~K.}\ \bibnamefont {Stephens}},\ }\bibfield  {title} {\bibinfo {title} {Effects of electrojet turbulence on a magnetosphere-ionosphere simulation of a geomagnetic storm},\ }\href {https://doi.org/10.1002/2016JA023700} {\bibfield  {journal} {\bibinfo  {journal} {Journal of Geophysical Research: Space Physics}\ }\textbf
  {\bibinfo {volume} {122}},\ \bibinfo {pages} {5008} (\bibinfo {year} {2017})}\BibitemShut {NoStop}%
\bibitem [{\citenamefont {{St-Maurice}}\ and\ \citenamefont {Goodwin}(2021)}]{st-mauriceRevisitingBehaviorERegion2021}%
  \BibitemOpen
  \bibfield  {author} {\bibinfo {author} {\bibfnamefont {J.-P.}\ \bibnamefont {{St-Maurice}}}\ and\ \bibinfo {author} {\bibfnamefont {L.}~\bibnamefont {Goodwin}},\ }\bibfield  {title} {\bibinfo {title} {Revisiting the {{Behavior}} of the {{E-Region Electron Temperature During Strong Electric Field Events}} at {{High Latitudes}}},\ }\href {https://doi.org/10.1029/2020JA028288} {\bibfield  {journal} {\bibinfo  {journal} {Journal of Geophysical Research: Space Physics}\ }\textbf {\bibinfo {volume} {126}},\ \bibinfo {pages} {2020JA028288} (\bibinfo {year} {2021})}\BibitemShut {NoStop}%
\bibitem [{\citenamefont {Pr{\"o}lss}(2004)}]{prolssAbsorptionDissipationSolar2004a}%
  \BibitemOpen
  \bibfield  {author} {\bibinfo {author} {\bibfnamefont {G.~W.}\ \bibnamefont {Pr{\"o}lss}},\ }\bibfield  {title} {\bibinfo {title} {Absorption and {{Dissipation}} of {{Solar Wind Energy}}},\ }in\ \href {https://doi.org/10.1007/978-3-642-97123-5_7} {\emph {\bibinfo {booktitle} {Physics of the {{Earth}}'s {{Space Environment}}: {{An Introduction}}}}},\ \bibinfo {editor} {edited by\ \bibinfo {editor} {\bibfnamefont {G.~W.}\ \bibnamefont {Pr{\"o}lss}}}\ (\bibinfo  {publisher} {Springer},\ \bibinfo {address} {Berlin, Heidelberg},\ \bibinfo {year} {2004})\ pp.\ \bibinfo {pages} {349--399}\BibitemShut {NoStop}%
\bibitem [{\citenamefont {Hysell}(2015)}]{hysellRadarAurora2015}%
  \BibitemOpen
  \bibfield  {author} {\bibinfo {author} {\bibfnamefont {D.~L.}\ \bibnamefont {Hysell}},\ }\bibfield  {title} {\bibinfo {title} {The {{Radar Aurora}}},\ }in\ \href {https://doi.org/10.1002/9781118978719.ch14} {\emph {\bibinfo {booktitle} {Auroral {{Dynamics}} and {{Space Weather}}}}}\ (\bibinfo  {publisher} {American Geophysical Union (AGU)},\ \bibinfo {year} {2015})\ Chap.~\bibinfo {chapter} {14}, pp.\ \bibinfo {pages} {191--209}\BibitemShut {NoStop}%
\bibitem [{\citenamefont {Buneman}(1963)}]{bunemanExcitationFieldAligned1963}%
  \BibitemOpen
  \bibfield  {author} {\bibinfo {author} {\bibfnamefont {O.}~\bibnamefont {Buneman}},\ }\bibfield  {title} {\bibinfo {title} {Excitation of {{Field Aligned Sound Waves}} by {{Electron Streams}}},\ }\href {https://doi.org/10.1103/PhysRevLett.10.285} {\bibfield  {journal} {\bibinfo  {journal} {Physical Review Letters}\ }\textbf {\bibinfo {volume} {10}},\ \bibinfo {pages} {285} (\bibinfo {year} {1963})}\BibitemShut {NoStop}%
\bibitem [{\citenamefont {Farley}(1963)}]{farleyPlasmaInstabilityResulting1963}%
  \BibitemOpen
  \bibfield  {author} {\bibinfo {author} {\bibfnamefont {D.~T.}\ \bibnamefont {Farley}},\ }\bibfield  {title} {\bibinfo {title} {A plasma instability resulting in field-aligned irregularities in the ionosphere},\ }\href {https://doi.org/10.1029/JZ068i022p06083} {\bibfield  {journal} {\bibinfo  {journal} {Journal of Geophysical Research (1896-1977)}\ }\textbf {\bibinfo {volume} {68}},\ \bibinfo {pages} {6083} (\bibinfo {year} {1963})}\BibitemShut {NoStop}%
\bibitem [{\citenamefont {Fejer}\ and\ \citenamefont {Kelley}(1980)}]{fejerIonosphericIrregularities1980}%
  \BibitemOpen
  \bibfield  {author} {\bibinfo {author} {\bibfnamefont {B.~G.}\ \bibnamefont {Fejer}}\ and\ \bibinfo {author} {\bibfnamefont {M.~C.}\ \bibnamefont {Kelley}},\ }\bibfield  {title} {\bibinfo {title} {Ionospheric irregularities},\ }\href {https://doi.org/10.1029/RG018i002p00401} {\bibfield  {journal} {\bibinfo  {journal} {Reviews of Geophysics}\ }\textbf {\bibinfo {volume} {18}},\ \bibinfo {pages} {401} (\bibinfo {year} {1980})}\BibitemShut {NoStop}%
\bibitem [{\citenamefont {Bowles}\ \emph {et~al.}(1963)\citenamefont {Bowles}, \citenamefont {Balsley},\ and\ \citenamefont {Cohen}}]{bowles_field-aligned_1963}%
  \BibitemOpen
  \bibfield  {author} {\bibinfo {author} {\bibfnamefont {K.~L.}\ \bibnamefont {Bowles}}, \bibinfo {author} {\bibfnamefont {B.~B.}\ \bibnamefont {Balsley}},\ and\ \bibinfo {author} {\bibfnamefont {R.}~\bibnamefont {Cohen}},\ }\bibfield  {title} {\bibinfo {title} {Field-aligned {{E-region}} irregularities identified with acoustic plasma waves},\ }\href {https://doi.org/10.1029/JZ068i009p02485} {\bibfield  {journal} {\bibinfo  {journal} {Journal of Geophysical Research (1896-1977)}\ }\textbf {\bibinfo {volume} {68}},\ \bibinfo {pages} {2485} (\bibinfo {year} {1963})}\BibitemShut {NoStop}%
\bibitem [{\citenamefont {Drexler}\ \emph {et~al.}(2002)\citenamefont {Drexler}, \citenamefont {{St. -Maurice}}, \citenamefont {Chen},\ and\ \citenamefont {Moorcroft}}]{drexlerNewInsightsNonlocal2002}%
  \BibitemOpen
  \bibfield  {author} {\bibinfo {author} {\bibfnamefont {J.}~\bibnamefont {Drexler}}, \bibinfo {author} {\bibfnamefont {J.-P.}\ \bibnamefont {{St. -Maurice}}}, \bibinfo {author} {\bibfnamefont {D.}~\bibnamefont {Chen}},\ and\ \bibinfo {author} {\bibfnamefont {D.~R.}\ \bibnamefont {Moorcroft}},\ }\bibfield  {title} {\bibinfo {title} {New insights from a nonlocal generalization of the {{Farley-Buneman}} instability problem at high latitudes},\ }\href {https://doi.org/10.5194/angeo-20-2003-2002} {\bibfield  {journal} {\bibinfo  {journal} {Annales Geophysicae}\ }\textbf {\bibinfo {volume} {20}},\ \bibinfo {pages} {2003} (\bibinfo {year} {2002})}\BibitemShut {NoStop}%
\bibitem [{\citenamefont {Huyghebaert}\ \emph {et~al.}(2021)\citenamefont {Huyghebaert}, \citenamefont {{St.-Maurice}}, \citenamefont {McWilliams}, \citenamefont {Hussey}, \citenamefont {Howarth}, \citenamefont {Rutledge},\ and\ \citenamefont {Erion}}]{huyghebaertPropertiesICEBEARERegion2021}%
  \BibitemOpen
  \bibfield  {author} {\bibinfo {author} {\bibfnamefont {D.}~\bibnamefont {Huyghebaert}}, \bibinfo {author} {\bibfnamefont {J.-P.}\ \bibnamefont {{St.-Maurice}}}, \bibinfo {author} {\bibfnamefont {K.}~\bibnamefont {McWilliams}}, \bibinfo {author} {\bibfnamefont {G.}~\bibnamefont {Hussey}}, \bibinfo {author} {\bibfnamefont {A.~D.}\ \bibnamefont {Howarth}}, \bibinfo {author} {\bibfnamefont {P.}~\bibnamefont {Rutledge}},\ and\ \bibinfo {author} {\bibfnamefont {S.}~\bibnamefont {Erion}},\ }\bibfield  {title} {\bibinfo {title} {The {{Properties}} of {{ICEBEAR E-Region Coherent Radar Echoes}} in the {{Presence}} of {{Near Infrared Auroral Emissions}}, as {{Measured}} by the {{Swarm-E Fast Auroral Imager}}},\ }\href {https://doi.org/10.1029/2021JA029857} {\bibfield  {journal} {\bibinfo  {journal} {Journal of Geophysical Research: Space Physics}\ }\textbf {\bibinfo {volume} {126}},\ \bibinfo {pages} {e2021JA029857} (\bibinfo {year} {2021})}\BibitemShut {NoStop}%
\bibitem [{\citenamefont {Ivarsen}\ \emph {et~al.}(2024{\natexlab{b}})\citenamefont {Ivarsen}, \citenamefont {Huyghebaert}, \citenamefont {Gilles}, \citenamefont {{St-Maurice}}, \citenamefont {Themens}, \citenamefont {Oppenheim}, \citenamefont {Gustavsson}, \citenamefont {Billett}, \citenamefont {Pitzel}, \citenamefont {Donovan},\ and\ \citenamefont {Hussey}}]{ivarsenTurbulenceAuroras2024}%
  \BibitemOpen
  \bibfield  {author} {\bibinfo {author} {\bibfnamefont {M.~F.}\ \bibnamefont {Ivarsen}}, \bibinfo {author} {\bibfnamefont {D.~R.}\ \bibnamefont {Huyghebaert}}, \bibinfo {author} {\bibfnamefont {D.~M.}\ \bibnamefont {Gilles}}, \bibinfo {author} {\bibfnamefont {J.~P.}\ \bibnamefont {{St-Maurice}}}, \bibinfo {author} {\bibfnamefont {D.~R.}\ \bibnamefont {Themens}}, \bibinfo {author} {\bibfnamefont {M.~M.}\ \bibnamefont {Oppenheim}}, \bibinfo {author} {\bibfnamefont {B.}~\bibnamefont {Gustavsson}}, \bibinfo {author} {\bibfnamefont {D.}~\bibnamefont {Billett}}, \bibinfo {author} {\bibfnamefont {B.}~\bibnamefont {Pitzel}}, \bibinfo {author} {\bibfnamefont {E.}~\bibnamefont {Donovan}},\ and\ \bibinfo {author} {\bibfnamefont {G.~C.}\ \bibnamefont {Hussey}},\ }\bibfield  {title} {\bibinfo {title} {Turbulence around {{Auroral Arcs}}},\ }\href@noop {} {\bibfield  {journal} {\bibinfo  {journal} {Currently being revised for publication in JGR: Space Physics}\ } (\bibinfo {year} {2024}{\natexlab{b}})}\BibitemShut
  {NoStop}%
\bibitem [{\citenamefont {Hall}\ \emph {et~al.}(1990)\citenamefont {Hall}, \citenamefont {Moorcroft}, \citenamefont {Cogger},\ and\ \citenamefont {Andre}}]{hallSpatialRelationshipLarge1990}%
  \BibitemOpen
  \bibfield  {author} {\bibinfo {author} {\bibfnamefont {G.}~\bibnamefont {Hall}}, \bibinfo {author} {\bibfnamefont {D.~R.}\ \bibnamefont {Moorcroft}}, \bibinfo {author} {\bibfnamefont {L.~L.}\ \bibnamefont {Cogger}},\ and\ \bibinfo {author} {\bibfnamefont {D.}~\bibnamefont {Andre}},\ }\bibfield  {title} {\bibinfo {title} {Spatial relationship between large aspect angle {{VHF}} radio aurora and 557.7-nm emissions: {{Evidence}} for refraction},\ }\href {https://doi.org/10.1029/JA095iA09p15281} {\bibfield  {journal} {\bibinfo  {journal} {Journal of Geophysical Research: Space Physics}\ }\textbf {\bibinfo {volume} {95}},\ \bibinfo {pages} {15281} (\bibinfo {year} {1990})}\BibitemShut {NoStop}%
\bibitem [{\citenamefont {Bahcivan}\ \emph {et~al.}(2006)\citenamefont {Bahcivan}, \citenamefont {Hysell}, \citenamefont {Lummerzheim}, \citenamefont {Larsen},\ and\ \citenamefont {Pfaff}}]{bahcivanObservationsColocatedOptical2006}%
  \BibitemOpen
  \bibfield  {author} {\bibinfo {author} {\bibfnamefont {H.}~\bibnamefont {Bahcivan}}, \bibinfo {author} {\bibfnamefont {D.~L.}\ \bibnamefont {Hysell}}, \bibinfo {author} {\bibfnamefont {D.}~\bibnamefont {Lummerzheim}}, \bibinfo {author} {\bibfnamefont {M.~F.}\ \bibnamefont {Larsen}},\ and\ \bibinfo {author} {\bibfnamefont {R.~F.}\ \bibnamefont {Pfaff}},\ }\bibfield  {title} {\bibinfo {title} {Observations of colocated optical and radar aurora},\ }\bibfield  {journal} {\bibinfo  {journal} {Journal of Geophysical Research: Space Physics}\ }\textbf {\bibinfo {volume} {111}},\ \href {https://doi.org/10.1029/2006JA011923} {10.1029/2006JA011923} (\bibinfo {year} {2006})\BibitemShut {NoStop}%
\bibitem [{\citenamefont {Hysell}\ \emph {et~al.}(2012)\citenamefont {Hysell}, \citenamefont {Miceli}, \citenamefont {Munk}, \citenamefont {Hampton}, \citenamefont {Heinselman}, \citenamefont {Nicolls}, \citenamefont {Powell}, \citenamefont {Lynch},\ and\ \citenamefont {Lessard}}]{hysellComparingVHFCoherent2012}%
  \BibitemOpen
  \bibfield  {author} {\bibinfo {author} {\bibfnamefont {D.~L.}\ \bibnamefont {Hysell}}, \bibinfo {author} {\bibfnamefont {R.}~\bibnamefont {Miceli}}, \bibinfo {author} {\bibfnamefont {J.}~\bibnamefont {Munk}}, \bibinfo {author} {\bibfnamefont {D.}~\bibnamefont {Hampton}}, \bibinfo {author} {\bibfnamefont {C.}~\bibnamefont {Heinselman}}, \bibinfo {author} {\bibfnamefont {M.}~\bibnamefont {Nicolls}}, \bibinfo {author} {\bibfnamefont {S.}~\bibnamefont {Powell}}, \bibinfo {author} {\bibfnamefont {K.}~\bibnamefont {Lynch}},\ and\ \bibinfo {author} {\bibfnamefont {M.}~\bibnamefont {Lessard}},\ }\bibfield  {title} {\bibinfo {title} {Comparing {{VHF}} coherent scatter from the radar aurora with incoherent scatter and all-sky auroral imagery},\ }\bibfield  {journal} {\bibinfo  {journal} {Journal of Geophysical Research: Space Physics}\ }\textbf {\bibinfo {volume} {117}},\ \href {https://doi.org/10.1029/2012JA018010} {10.1029/2012JA018010} (\bibinfo {year} {2012})\BibitemShut {NoStop}%
\bibitem [{\citenamefont {Borovsky}\ \emph {et~al.}(2019)\citenamefont {Borovsky}, \citenamefont {Birn}, \citenamefont {Echim}, \citenamefont {Fujita}, \citenamefont {Lysak}, \citenamefont {Knudsen}, \citenamefont {Marghitu}, \citenamefont {Otto}, \citenamefont {Watanabe},\ and\ \citenamefont {Tanaka}}]{borovskyQuiescentDiscreteAuroral2019}%
  \BibitemOpen
  \bibfield  {author} {\bibinfo {author} {\bibfnamefont {J.~E.}\ \bibnamefont {Borovsky}}, \bibinfo {author} {\bibfnamefont {J.}~\bibnamefont {Birn}}, \bibinfo {author} {\bibfnamefont {M.~M.}\ \bibnamefont {Echim}}, \bibinfo {author} {\bibfnamefont {S.}~\bibnamefont {Fujita}}, \bibinfo {author} {\bibfnamefont {R.~L.}\ \bibnamefont {Lysak}}, \bibinfo {author} {\bibfnamefont {D.~J.}\ \bibnamefont {Knudsen}}, \bibinfo {author} {\bibfnamefont {O.}~\bibnamefont {Marghitu}}, \bibinfo {author} {\bibfnamefont {A.}~\bibnamefont {Otto}}, \bibinfo {author} {\bibfnamefont {T.-H.}\ \bibnamefont {Watanabe}},\ and\ \bibinfo {author} {\bibfnamefont {T.}~\bibnamefont {Tanaka}},\ }\bibfield  {title} {\bibinfo {title} {Quiescent {{Discrete Auroral Arcs}}: {{A Review}} of {{Magnetospheric Generator Mechanisms}}},\ }\href {https://doi.org/10.1007/s11214-019-0619-5} {\bibfield  {journal} {\bibinfo  {journal} {Space Science Reviews}\ }\textbf {\bibinfo {volume} {216}},\ \bibinfo {pages} {1} (\bibinfo {year} {2019})}\BibitemShut
  {NoStop}%
\bibitem [{\citenamefont {Chaston}\ \emph {et~al.}(2010)\citenamefont {Chaston}, \citenamefont {Seki}, \citenamefont {Sakanoi}, \citenamefont {Asamura},\ and\ \citenamefont {Hirahara}}]{chastonMotionAurorae2010}%
  \BibitemOpen
  \bibfield  {author} {\bibinfo {author} {\bibfnamefont {{\relax Christopher}.~C.}\ \bibnamefont {Chaston}}, \bibinfo {author} {\bibfnamefont {K.}~\bibnamefont {Seki}}, \bibinfo {author} {\bibfnamefont {T.}~\bibnamefont {Sakanoi}}, \bibinfo {author} {\bibfnamefont {K.}~\bibnamefont {Asamura}},\ and\ \bibinfo {author} {\bibfnamefont {M.}~\bibnamefont {Hirahara}},\ }\bibfield  {title} {\bibinfo {title} {Motion of aurorae},\ }\bibfield  {journal} {\bibinfo  {journal} {Geophysical Research Letters}\ }\textbf {\bibinfo {volume} {37}},\ \href {https://doi.org/10.1029/2009GL042117} {10.1029/2009GL042117} (\bibinfo {year} {2010})\BibitemShut {NoStop}%
\bibitem [{\citenamefont {Ivarsen}\ \emph {et~al.}(2024{\natexlab{c}})\citenamefont {Ivarsen}, \citenamefont {{St-Maurice}}, \citenamefont {Huyghebaert}, \citenamefont {Gillies}, \citenamefont {Lind}, \citenamefont {Pitzel},\ and\ \citenamefont {Hussey}}]{ivarsen_deriving_2024}%
  \BibitemOpen
  \bibfield  {author} {\bibinfo {author} {\bibfnamefont {M.~F.}\ \bibnamefont {Ivarsen}}, \bibinfo {author} {\bibfnamefont {J.-P.}\ \bibnamefont {{St-Maurice}}}, \bibinfo {author} {\bibfnamefont {D.}~\bibnamefont {Huyghebaert}}, \bibinfo {author} {\bibfnamefont {M.}~\bibnamefont {Gillies}}, \bibinfo {author} {\bibfnamefont {F.~D.}\ \bibnamefont {Lind}}, \bibinfo {author} {\bibfnamefont {B.}~\bibnamefont {Pitzel}},\ and\ \bibinfo {author} {\bibfnamefont {G.~C.}\ \bibnamefont {Hussey}},\ }\href@noop {} {\bibinfo {title} {Deriving the {{Ionospheric Electric Field}} from the {{Bulk Motion}} of {{Radar Aurora}} in the {{E-region}}}} (\bibinfo {year} {2024}{\natexlab{c}})\BibitemShut {NoStop}%
\bibitem [{\citenamefont {Huyghebaert}\ \emph {et~al.}(2019)\citenamefont {Huyghebaert}, \citenamefont {Hussey}, \citenamefont {Vierinen}, \citenamefont {McWilliams},\ and\ \citenamefont {{St-Maurice}}}]{huyghebaertICEBEARAlldigitalBistatic2019}%
  \BibitemOpen
  \bibfield  {author} {\bibinfo {author} {\bibfnamefont {D.}~\bibnamefont {Huyghebaert}}, \bibinfo {author} {\bibfnamefont {G.}~\bibnamefont {Hussey}}, \bibinfo {author} {\bibfnamefont {J.}~\bibnamefont {Vierinen}}, \bibinfo {author} {\bibfnamefont {K.}~\bibnamefont {McWilliams}},\ and\ \bibinfo {author} {\bibfnamefont {J.-P.}\ \bibnamefont {{St-Maurice}}},\ }\bibfield  {title} {\bibinfo {title} {{{ICEBEAR}}: {{An}} all-digital bistatic coded continuous-wave radar for studies of the {{E}} region of the ionosphere},\ }\href {https://doi.org/10.1029/2018RS006747} {\bibfield  {journal} {\bibinfo  {journal} {Radio Science}\ }\textbf {\bibinfo {volume} {54}},\ \bibinfo {pages} {349} (\bibinfo {year} {2019})}\BibitemShut {NoStop}%
\bibitem [{\citenamefont {Ester}\ \emph {et~al.}(1996)\citenamefont {Ester}, \citenamefont {Kriegel}, \citenamefont {Sander},\ and\ \citenamefont {Xu}}]{esterDensitybasedAlgorithmDiscovering1996}%
  \BibitemOpen
  \bibfield  {author} {\bibinfo {author} {\bibfnamefont {M.}~\bibnamefont {Ester}}, \bibinfo {author} {\bibfnamefont {H.-P.}\ \bibnamefont {Kriegel}}, \bibinfo {author} {\bibfnamefont {J.}~\bibnamefont {Sander}},\ and\ \bibinfo {author} {\bibfnamefont {X.}~\bibnamefont {Xu}},\ }\bibfield  {title} {\bibinfo {title} {A density-based algorithm for discovering clusters in large spatial databases with noise},\ }in\ \href@noop {} {\emph {\bibinfo {booktitle} {Kdd}}},\ Vol.~\bibinfo {volume} {96}\ (\bibinfo {year} {1996})\ pp.\ \bibinfo {pages} {226--231}\BibitemShut {NoStop}%
\bibitem [{\citenamefont {Khan}\ \emph {et~al.}(2014)\citenamefont {Khan}, \citenamefont {Rehman}, \citenamefont {Aziz}, \citenamefont {Fong},\ and\ \citenamefont {Sarasvady}}]{khanDBSCANPresentFuture2014}%
  \BibitemOpen
  \bibfield  {author} {\bibinfo {author} {\bibfnamefont {K.}~\bibnamefont {Khan}}, \bibinfo {author} {\bibfnamefont {S.~U.}\ \bibnamefont {Rehman}}, \bibinfo {author} {\bibfnamefont {K.}~\bibnamefont {Aziz}}, \bibinfo {author} {\bibfnamefont {S.}~\bibnamefont {Fong}},\ and\ \bibinfo {author} {\bibfnamefont {S.}~\bibnamefont {Sarasvady}},\ }\bibfield  {title} {\bibinfo {title} {{{DBSCAN}}: {{Past}}, present and future},\ }in\ \href {https://doi.org/10.1109/ICADIWT.2014.6814687} {\emph {\bibinfo {booktitle} {The {{Fifth International Conference}} on the {{Applications}} of {{Digital Information}} and {{Web Technologies}} ({{ICADIWT}} 2014)}}}\ (\bibinfo {year} {2014})\ pp.\ \bibinfo {pages} {232--238}\BibitemShut {NoStop}%
\bibitem [{\citenamefont {Sander}(2010)}]{sander_density-based_2010}%
  \BibitemOpen
  \bibfield  {author} {\bibinfo {author} {\bibfnamefont {J.}~\bibnamefont {Sander}},\ }\bibfield  {title} {\bibinfo {title} {Density-{{Based Clustering}}},\ }in\ \href {https://doi.org/10.1007/978-0-387-30164-8_211} {\emph {\bibinfo {booktitle} {Encyclopedia of {{Machine Learning}}}}},\ \bibinfo {editor} {edited by\ \bibinfo {editor} {\bibfnamefont {C.}~\bibnamefont {Sammut}}\ and\ \bibinfo {editor} {\bibfnamefont {G.~I.}\ \bibnamefont {Webb}}}\ (\bibinfo  {publisher} {Springer US},\ \bibinfo {address} {Boston, MA},\ \bibinfo {year} {2010})\ pp.\ \bibinfo {pages} {270--273}\BibitemShut {NoStop}%
\bibitem [{\citenamefont {Barron}\ \emph {et~al.}(1994)\citenamefont {Barron}, \citenamefont {Fleet},\ and\ \citenamefont {Beauchemin}}]{barronPerformanceOpticalFlow1994}%
  \BibitemOpen
  \bibfield  {author} {\bibinfo {author} {\bibfnamefont {J.~L.}\ \bibnamefont {Barron}}, \bibinfo {author} {\bibfnamefont {D.~J.}\ \bibnamefont {Fleet}},\ and\ \bibinfo {author} {\bibfnamefont {S.~S.}\ \bibnamefont {Beauchemin}},\ }\bibfield  {title} {\bibinfo {title} {Performance of optical flow techniques},\ }\href {https://doi.org/10.1007/BF01420984} {\bibfield  {journal} {\bibinfo  {journal} {International Journal of Computer Vision}\ }\textbf {\bibinfo {volume} {12}},\ \bibinfo {pages} {43} (\bibinfo {year} {1994})}\BibitemShut {NoStop}%
\bibitem [{\citenamefont {Maynard}\ \emph {et~al.}(1973)\citenamefont {Maynard}, \citenamefont {Bahnsen}, \citenamefont {Christophersen}, \citenamefont {Egeland},\ and\ \citenamefont {Lundin}}]{maynardExampleAnticorrelationAuroral1973}%
  \BibitemOpen
  \bibfield  {author} {\bibinfo {author} {\bibfnamefont {N.~C.}\ \bibnamefont {Maynard}}, \bibinfo {author} {\bibfnamefont {A.}~\bibnamefont {Bahnsen}}, \bibinfo {author} {\bibfnamefont {P.}~\bibnamefont {Christophersen}}, \bibinfo {author} {\bibfnamefont {A.}~\bibnamefont {Egeland}},\ and\ \bibinfo {author} {\bibfnamefont {R.}~\bibnamefont {Lundin}},\ }\bibfield  {title} {\bibinfo {title} {An example of anticorrelation of auroral particles and electric fields},\ }\bibfield  {journal} {\bibinfo  {journal} {J. Geophys. Res., v. 78, no. 19, pp. 3976-3980}\ }\href {https://doi.org/10.1029/JA078i019p03976} {10.1029/JA078i019p03976} (\bibinfo {year} {1973})\BibitemShut {NoStop}%
\bibitem [{\citenamefont {Nielsen}\ and\ \citenamefont {Schlegel}(1983)}]{nielsen_first_1983}%
  \BibitemOpen
  \bibfield  {author} {\bibinfo {author} {\bibfnamefont {E.}~\bibnamefont {Nielsen}}\ and\ \bibinfo {author} {\bibfnamefont {K.}~\bibnamefont {Schlegel}},\ }\bibfield  {title} {\bibinfo {title} {A first comparison of {{STARE}} and {{EISCAT}} electron drift velocity measurements},\ }\href {https://doi.org/10.1029/JA088iA07p05745} {\bibfield  {journal} {\bibinfo  {journal} {Journal of Geophysical Research}\ }\textbf {\bibinfo {volume} {88}},\ \bibinfo {pages} {5745} (\bibinfo {year} {1983})}\BibitemShut {NoStop}%
\bibitem [{\citenamefont {Foster}\ and\ \citenamefont {Erickson}(2000)}]{fosterSimultaneousObservationsEregion2000}%
  \BibitemOpen
  \bibfield  {author} {\bibinfo {author} {\bibfnamefont {J.~C.}\ \bibnamefont {Foster}}\ and\ \bibinfo {author} {\bibfnamefont {P.~J.}\ \bibnamefont {Erickson}},\ }\bibfield  {title} {\bibinfo {title} {Simultaneous observations of {{E-region}} coherent backscatter and electric field amplitude at {{F-region}} heights with the {{Millstone Hill UHF Radar}}},\ }\href {https://doi.org/10.1029/2000GL000042} {\bibfield  {journal} {\bibinfo  {journal} {Geophysical Research Letters}\ }\textbf {\bibinfo {volume} {27}},\ \bibinfo {pages} {3177} (\bibinfo {year} {2000})}\BibitemShut {NoStop}%
\bibitem [{\citenamefont {Koustov}\ \emph {et~al.}(2005)\citenamefont {Koustov}, \citenamefont {Danskin}, \citenamefont {Makarevitch},\ and\ \citenamefont {Gorin}}]{koustovRelationshipVelocityEregion2005}%
  \BibitemOpen
  \bibfield  {author} {\bibinfo {author} {\bibfnamefont {A.~V.}\ \bibnamefont {Koustov}}, \bibinfo {author} {\bibfnamefont {D.~W.}\ \bibnamefont {Danskin}}, \bibinfo {author} {\bibfnamefont {R.~A.}\ \bibnamefont {Makarevitch}},\ and\ \bibinfo {author} {\bibfnamefont {J.~D.}\ \bibnamefont {Gorin}},\ }\bibfield  {title} {\bibinfo {title} {On the relationship between the velocity of {{E-region HF}} echoes and {{{\emph{E}}}}x{{{\emph{B}}}} plasma drift},\ }\href {https://doi.org/10.5194/angeo-23-371-2005} {\bibfield  {journal} {\bibinfo  {journal} {Annales Geophysicae}\ }\textbf {\bibinfo {volume} {23}},\ \bibinfo {pages} {371} (\bibinfo {year} {2005})}\BibitemShut {NoStop}%
\bibitem [{\citenamefont {Marklund}(2009)}]{marklund_electric_2009}%
  \BibitemOpen
  \bibfield  {author} {\bibinfo {author} {\bibfnamefont {G.}~\bibnamefont {Marklund}},\ }\bibfield  {title} {\bibinfo {title} {Electric {{Fields}} and {{Plasma Processes}} in the {{Auroral Downward Current Region}}, {{Below}}, {{Within}}, and {{Above}} the {{Acceleration Region}}},\ }\href {https://doi.org/10.1007/s11214-008-9373-9} {\bibfield  {journal} {\bibinfo  {journal} {Space Science Reviews}\ }\textbf {\bibinfo {volume} {142}},\ \bibinfo {pages} {1} (\bibinfo {year} {2009})}\BibitemShut {NoStop}%
\bibitem [{\citenamefont {Hosokawa}\ \emph {et~al.}(2013)\citenamefont {Hosokawa}, \citenamefont {Milan}, \citenamefont {Lester}, \citenamefont {Kadokura}, \citenamefont {Sato},\ and\ \citenamefont {Bjornsson}}]{hosokawaLargeFlowShears2013}%
  \BibitemOpen
  \bibfield  {author} {\bibinfo {author} {\bibfnamefont {K.}~\bibnamefont {Hosokawa}}, \bibinfo {author} {\bibfnamefont {S.~E.}\ \bibnamefont {Milan}}, \bibinfo {author} {\bibfnamefont {M.}~\bibnamefont {Lester}}, \bibinfo {author} {\bibfnamefont {A.}~\bibnamefont {Kadokura}}, \bibinfo {author} {\bibfnamefont {N.}~\bibnamefont {Sato}},\ and\ \bibinfo {author} {\bibfnamefont {G.}~\bibnamefont {Bjornsson}},\ }\bibfield  {title} {\bibinfo {title} {Large flow shears around auroral beads at substorm onset},\ }\href@noop {} {\bibfield  {journal} {\bibinfo  {journal} {Geophysical Research Letters}\ }\textbf {\bibinfo {volume} {40}},\ \bibinfo {pages} {4987} (\bibinfo {year} {2013})}\BibitemShut {NoStop}%
\bibitem [{\citenamefont {{Gallardo-Lacourt}}\ \emph {et~al.}(2014)\citenamefont {{Gallardo-Lacourt}}, \citenamefont {Nishimura}, \citenamefont {Lyons}, \citenamefont {Ruohoniemi}, \citenamefont {Donovan}, \citenamefont {Angelopoulos}, \citenamefont {McWilliams},\ and\ \citenamefont {Nishitani}}]{gallardo-lacourt_ionospheric_2014}%
  \BibitemOpen
  \bibfield  {author} {\bibinfo {author} {\bibfnamefont {B.}~\bibnamefont {{Gallardo-Lacourt}}}, \bibinfo {author} {\bibfnamefont {Y.}~\bibnamefont {Nishimura}}, \bibinfo {author} {\bibfnamefont {L.~R.}\ \bibnamefont {Lyons}}, \bibinfo {author} {\bibfnamefont {J.~M.}\ \bibnamefont {Ruohoniemi}}, \bibinfo {author} {\bibfnamefont {E.}~\bibnamefont {Donovan}}, \bibinfo {author} {\bibfnamefont {V.}~\bibnamefont {Angelopoulos}}, \bibinfo {author} {\bibfnamefont {K.~A.}\ \bibnamefont {McWilliams}},\ and\ \bibinfo {author} {\bibfnamefont {N.}~\bibnamefont {Nishitani}},\ }\bibfield  {title} {\bibinfo {title} {Ionospheric flow structures associated with auroral beading at substorm auroral onset},\ }\href {https://doi.org/10.1002/2014JA020298} {\bibfield  {journal} {\bibinfo  {journal} {Journal of Geophysical Research: Space Physics}\ }\textbf {\bibinfo {volume} {119}},\ \bibinfo {pages} {9150} (\bibinfo {year} {2014})}\BibitemShut {NoStop}%
\bibitem [{\citenamefont {Dubyagin}\ \emph {et~al.}(2003)\citenamefont {Dubyagin}, \citenamefont {Sergeev}, \citenamefont {Carlson}, \citenamefont {Marple}, \citenamefont {Pulkkinen},\ and\ \citenamefont {Yahnin}}]{dubyagin_evidence_2003}%
  \BibitemOpen
  \bibfield  {author} {\bibinfo {author} {\bibfnamefont {S.~V.}\ \bibnamefont {Dubyagin}}, \bibinfo {author} {\bibfnamefont {V.~A.}\ \bibnamefont {Sergeev}}, \bibinfo {author} {\bibfnamefont {C.~W.}\ \bibnamefont {Carlson}}, \bibinfo {author} {\bibfnamefont {S.~R.}\ \bibnamefont {Marple}}, \bibinfo {author} {\bibfnamefont {T.~I.}\ \bibnamefont {Pulkkinen}},\ and\ \bibinfo {author} {\bibfnamefont {A.~G.}\ \bibnamefont {Yahnin}},\ }\bibfield  {title} {\bibinfo {title} {Evidence of near-{{Earth}} breakup location},\ }\bibfield  {journal} {\bibinfo  {journal} {Geophysical Research Letters}\ }\textbf {\bibinfo {volume} {30}},\ \href {https://doi.org/10.1029/2002GL016569} {10.1029/2002GL016569} (\bibinfo {year} {2003})\BibitemShut {NoStop}%
\bibitem [{\citenamefont {Voronkov}\ \emph {et~al.}(1997)\citenamefont {Voronkov}, \citenamefont {Rankin}, \citenamefont {Frycz}, \citenamefont {Tikhonchuk},\ and\ \citenamefont {Samson}}]{voronkov_coupling_1997}%
  \BibitemOpen
  \bibfield  {author} {\bibinfo {author} {\bibfnamefont {I.}~\bibnamefont {Voronkov}}, \bibinfo {author} {\bibfnamefont {R.}~\bibnamefont {Rankin}}, \bibinfo {author} {\bibfnamefont {P.}~\bibnamefont {Frycz}}, \bibinfo {author} {\bibfnamefont {V.~T.}\ \bibnamefont {Tikhonchuk}},\ and\ \bibinfo {author} {\bibfnamefont {J.~C.}\ \bibnamefont {Samson}},\ }\bibfield  {title} {\bibinfo {title} {Coupling of shear flow and pressure gradient instabilities},\ }\href {https://doi.org/10.1029/97JA00386} {\bibfield  {journal} {\bibinfo  {journal} {Journal of Geophysical Research: Space Physics}\ }\textbf {\bibinfo {volume} {102}},\ \bibinfo {pages} {9639} (\bibinfo {year} {1997})}\BibitemShut {NoStop}%
\bibitem [{\citenamefont {Uritsky}\ \emph {et~al.}(2009)\citenamefont {Uritsky}, \citenamefont {Liang}, \citenamefont {Donovan}, \citenamefont {Spanswick}, \citenamefont {Knudsen}, \citenamefont {Liu}, \citenamefont {Bonnell},\ and\ \citenamefont {Glassmeier}}]{uritsky_longitudinally_2009}%
  \BibitemOpen
  \bibfield  {author} {\bibinfo {author} {\bibfnamefont {V.~M.}\ \bibnamefont {Uritsky}}, \bibinfo {author} {\bibfnamefont {J.}~\bibnamefont {Liang}}, \bibinfo {author} {\bibfnamefont {E.}~\bibnamefont {Donovan}}, \bibinfo {author} {\bibfnamefont {E.}~\bibnamefont {Spanswick}}, \bibinfo {author} {\bibfnamefont {D.}~\bibnamefont {Knudsen}}, \bibinfo {author} {\bibfnamefont {W.}~\bibnamefont {Liu}}, \bibinfo {author} {\bibfnamefont {J.}~\bibnamefont {Bonnell}},\ and\ \bibinfo {author} {\bibfnamefont {K.~H.}\ \bibnamefont {Glassmeier}},\ }\bibfield  {title} {\bibinfo {title} {Longitudinally propagating arc wave in the pre-onset optical aurora},\ }\bibfield  {journal} {\bibinfo  {journal} {Geophysical Research Letters}\ }\textbf {\bibinfo {volume} {36}},\ \href {https://doi.org/10.1029/2009GL040777} {10.1029/2009GL040777} (\bibinfo {year} {2009})\BibitemShut {NoStop}%
\bibitem [{\citenamefont {Ogasawara}\ \emph {et~al.}(2011)\citenamefont {Ogasawara}, \citenamefont {Kasaba}, \citenamefont {Nishimura}, \citenamefont {Hori}, \citenamefont {Takada}, \citenamefont {Miyashita}, \citenamefont {Angelopoulos}, \citenamefont {Mende},\ and\ \citenamefont {Bonnell}}]{ogasawara_azimuthal_2011}%
  \BibitemOpen
  \bibfield  {author} {\bibinfo {author} {\bibfnamefont {K.}~\bibnamefont {Ogasawara}}, \bibinfo {author} {\bibfnamefont {Y.}~\bibnamefont {Kasaba}}, \bibinfo {author} {\bibfnamefont {Y.}~\bibnamefont {Nishimura}}, \bibinfo {author} {\bibfnamefont {T.}~\bibnamefont {Hori}}, \bibinfo {author} {\bibfnamefont {T.}~\bibnamefont {Takada}}, \bibinfo {author} {\bibfnamefont {Y.}~\bibnamefont {Miyashita}}, \bibinfo {author} {\bibfnamefont {V.}~\bibnamefont {Angelopoulos}}, \bibinfo {author} {\bibfnamefont {S.~B.}\ \bibnamefont {Mende}},\ and\ \bibinfo {author} {\bibfnamefont {J.}~\bibnamefont {Bonnell}},\ }\bibfield  {title} {\bibinfo {title} {Azimuthal auroral expansion associated with fast flows in the near-{{Earth}} plasma sheet: {{Coordinated}} observations of the {{THEMIS}} all-sky imagers and multiple spacecraft},\ }\bibfield  {journal} {\bibinfo  {journal} {Journal of Geophysical Research: Space Physics}\ }\textbf {\bibinfo {volume} {116}},\ \href {https://doi.org/10.1029/2010JA016032} {10.1029/2010JA016032}
  (\bibinfo {year} {2011})\BibitemShut {NoStop}%
\bibitem [{\citenamefont {Miyashita}\ \emph {et~al.}(2021)\citenamefont {Miyashita}, \citenamefont {Chang}, \citenamefont {Miyoshi}, \citenamefont {Hori}, \citenamefont {Kadokura}, \citenamefont {Kasahara}, \citenamefont {Wang}, \citenamefont {Keika}, \citenamefont {Matsuoka}, \citenamefont {Tanaka}, \citenamefont {Kasahara}, \citenamefont {Teramoto}, \citenamefont {Jun}, \citenamefont {Asamura}, \citenamefont {Kazama}, \citenamefont {Tam}, \citenamefont {Wang}, \citenamefont {Yokota}, \citenamefont {Kumamoto}, \citenamefont {Tsuchiya}, \citenamefont {Shoji}, \citenamefont {Kurita}, \citenamefont {Imajo},\ and\ \citenamefont {Shinohara}}]{miyashitaMagneticFieldEnergetic2021}%
  \BibitemOpen
  \bibfield  {author} {\bibinfo {author} {\bibfnamefont {Y.}~\bibnamefont {Miyashita}}, \bibinfo {author} {\bibfnamefont {T.-F.}\ \bibnamefont {Chang}}, \bibinfo {author} {\bibfnamefont {Y.}~\bibnamefont {Miyoshi}}, \bibinfo {author} {\bibfnamefont {T.}~\bibnamefont {Hori}}, \bibinfo {author} {\bibfnamefont {A.}~\bibnamefont {Kadokura}}, \bibinfo {author} {\bibfnamefont {S.}~\bibnamefont {Kasahara}}, \bibinfo {author} {\bibfnamefont {S.-Y.}\ \bibnamefont {Wang}}, \bibinfo {author} {\bibfnamefont {K.}~\bibnamefont {Keika}}, \bibinfo {author} {\bibfnamefont {A.}~\bibnamefont {Matsuoka}}, \bibinfo {author} {\bibfnamefont {Y.}~\bibnamefont {Tanaka}}, \bibinfo {author} {\bibfnamefont {Y.}~\bibnamefont {Kasahara}}, \bibinfo {author} {\bibfnamefont {M.}~\bibnamefont {Teramoto}}, \bibinfo {author} {\bibfnamefont {C.-W.}\ \bibnamefont {Jun}}, \bibinfo {author} {\bibfnamefont {K.}~\bibnamefont {Asamura}}, \bibinfo {author} {\bibfnamefont {Y.}~\bibnamefont {Kazama}}, \bibinfo {author} {\bibfnamefont {S.~W.~Y.}\
  \bibnamefont {Tam}}, \bibinfo {author} {\bibfnamefont {B.-J.}\ \bibnamefont {Wang}}, \bibinfo {author} {\bibfnamefont {S.}~\bibnamefont {Yokota}}, \bibinfo {author} {\bibfnamefont {A.}~\bibnamefont {Kumamoto}}, \bibinfo {author} {\bibfnamefont {F.}~\bibnamefont {Tsuchiya}}, \bibinfo {author} {\bibfnamefont {M.}~\bibnamefont {Shoji}}, \bibinfo {author} {\bibfnamefont {S.}~\bibnamefont {Kurita}}, \bibinfo {author} {\bibfnamefont {S.}~\bibnamefont {Imajo}},\ and\ \bibinfo {author} {\bibfnamefont {I.}~\bibnamefont {Shinohara}},\ }\bibfield  {title} {\bibinfo {title} {Magnetic {{Field}} and {{Energetic Particle Flux Oscillations}} and {{High-Frequency Waves Deep}} in the {{Inner Magnetosphere During Substorm Dipolarization}}: {{ERG Observations}}},\ }\href {https://doi.org/10.1029/2020JA029095} {\bibfield  {journal} {\bibinfo  {journal} {Journal of Geophysical Research: Space Physics}\ }\textbf {\bibinfo {volume} {126}},\ \bibinfo {pages} {e2020JA029095} (\bibinfo {year} {2021})}\BibitemShut {NoStop}%
\bibitem [{\citenamefont {Thorne}\ \emph {et~al.}(2010)\citenamefont {Thorne}, \citenamefont {Ni}, \citenamefont {Tao}, \citenamefont {Horne},\ and\ \citenamefont {Meredith}}]{thorne_scattering_2010}%
  \BibitemOpen
  \bibfield  {author} {\bibinfo {author} {\bibfnamefont {R.~M.}\ \bibnamefont {Thorne}}, \bibinfo {author} {\bibfnamefont {B.}~\bibnamefont {Ni}}, \bibinfo {author} {\bibfnamefont {X.}~\bibnamefont {Tao}}, \bibinfo {author} {\bibfnamefont {R.~B.}\ \bibnamefont {Horne}},\ and\ \bibinfo {author} {\bibfnamefont {N.~P.}\ \bibnamefont {Meredith}},\ }\bibfield  {title} {\bibinfo {title} {Scattering by chorus waves as the dominant cause of diffuse auroral precipitation},\ }\href {https://doi.org/10.1038/nature09467} {\bibfield  {journal} {\bibinfo  {journal} {Nature}\ }\textbf {\bibinfo {volume} {467}},\ \bibinfo {pages} {943} (\bibinfo {year} {2010})}\BibitemShut {NoStop}%
\bibitem [{\citenamefont {Kasahara}\ \emph {et~al.}(2018)\citenamefont {Kasahara}, \citenamefont {Miyoshi}, \citenamefont {Yokota}, \citenamefont {Mitani}, \citenamefont {Kasahara}, \citenamefont {Matsuda}, \citenamefont {Kumamoto}, \citenamefont {Matsuoka}, \citenamefont {Kazama}, \citenamefont {Frey}, \citenamefont {Angelopoulos}, \citenamefont {Kurita}, \citenamefont {Keika}, \citenamefont {Seki},\ and\ \citenamefont {Shinohara}}]{kasaharaPulsatingAuroraElectron2018}%
  \BibitemOpen
  \bibfield  {author} {\bibinfo {author} {\bibfnamefont {S.}~\bibnamefont {Kasahara}}, \bibinfo {author} {\bibfnamefont {Y.}~\bibnamefont {Miyoshi}}, \bibinfo {author} {\bibfnamefont {S.}~\bibnamefont {Yokota}}, \bibinfo {author} {\bibfnamefont {T.}~\bibnamefont {Mitani}}, \bibinfo {author} {\bibfnamefont {Y.}~\bibnamefont {Kasahara}}, \bibinfo {author} {\bibfnamefont {S.}~\bibnamefont {Matsuda}}, \bibinfo {author} {\bibfnamefont {A.}~\bibnamefont {Kumamoto}}, \bibinfo {author} {\bibfnamefont {A.}~\bibnamefont {Matsuoka}}, \bibinfo {author} {\bibfnamefont {Y.}~\bibnamefont {Kazama}}, \bibinfo {author} {\bibfnamefont {H.~U.}\ \bibnamefont {Frey}}, \bibinfo {author} {\bibfnamefont {V.}~\bibnamefont {Angelopoulos}}, \bibinfo {author} {\bibfnamefont {S.}~\bibnamefont {Kurita}}, \bibinfo {author} {\bibfnamefont {K.}~\bibnamefont {Keika}}, \bibinfo {author} {\bibfnamefont {K.}~\bibnamefont {Seki}},\ and\ \bibinfo {author} {\bibfnamefont {I.}~\bibnamefont {Shinohara}},\ }\bibfield  {title} {\bibinfo {title}
  {Pulsating aurora from electron scattering by chorus waves},\ }\href {https://doi.org/10.1038/nature25505} {\bibfield  {journal} {\bibinfo  {journal} {Nature}\ }\textbf {\bibinfo {volume} {554}},\ \bibinfo {pages} {337} (\bibinfo {year} {2018})}\BibitemShut {NoStop}%
\bibitem [{\citenamefont {Hosokawa}\ \emph {et~al.}(2020)\citenamefont {Hosokawa}, \citenamefont {Miyoshi}, \citenamefont {Ozaki}, \citenamefont {Oyama}, \citenamefont {Ogawa}, \citenamefont {Kurita}, \citenamefont {Kasahara}, \citenamefont {Kasaba}, \citenamefont {Yagitani}, \citenamefont {Matsuda}, \citenamefont {Tsuchiya}, \citenamefont {Kumamoto}, \citenamefont {Kataoka}, \citenamefont {Shiokawa}, \citenamefont {Raita}, \citenamefont {Turunen}, \citenamefont {Takashima}, \citenamefont {Shinohara},\ and\ \citenamefont {Fujii}}]{hosokawaMultipleTimescaleBeats2020}%
  \BibitemOpen
  \bibfield  {author} {\bibinfo {author} {\bibfnamefont {K.}~\bibnamefont {Hosokawa}}, \bibinfo {author} {\bibfnamefont {Y.}~\bibnamefont {Miyoshi}}, \bibinfo {author} {\bibfnamefont {M.}~\bibnamefont {Ozaki}}, \bibinfo {author} {\bibfnamefont {S.-I.}\ \bibnamefont {Oyama}}, \bibinfo {author} {\bibfnamefont {Y.}~\bibnamefont {Ogawa}}, \bibinfo {author} {\bibfnamefont {S.}~\bibnamefont {Kurita}}, \bibinfo {author} {\bibfnamefont {Y.}~\bibnamefont {Kasahara}}, \bibinfo {author} {\bibfnamefont {Y.}~\bibnamefont {Kasaba}}, \bibinfo {author} {\bibfnamefont {S.}~\bibnamefont {Yagitani}}, \bibinfo {author} {\bibfnamefont {S.}~\bibnamefont {Matsuda}}, \bibinfo {author} {\bibfnamefont {F.}~\bibnamefont {Tsuchiya}}, \bibinfo {author} {\bibfnamefont {A.}~\bibnamefont {Kumamoto}}, \bibinfo {author} {\bibfnamefont {R.}~\bibnamefont {Kataoka}}, \bibinfo {author} {\bibfnamefont {K.}~\bibnamefont {Shiokawa}}, \bibinfo {author} {\bibfnamefont {T.}~\bibnamefont {Raita}}, \bibinfo {author} {\bibfnamefont {E.}~\bibnamefont
  {Turunen}}, \bibinfo {author} {\bibfnamefont {T.}~\bibnamefont {Takashima}}, \bibinfo {author} {\bibfnamefont {I.}~\bibnamefont {Shinohara}},\ and\ \bibinfo {author} {\bibfnamefont {R.}~\bibnamefont {Fujii}},\ }\bibfield  {title} {\bibinfo {title} {Multiple time-scale beats in aurora: Precise orchestration via magnetospheric chorus waves},\ }\href {https://doi.org/10.1038/s41598-020-59642-8} {\bibfield  {journal} {\bibinfo  {journal} {Scientific Reports}\ }\textbf {\bibinfo {volume} {10}},\ \bibinfo {pages} {3380} (\bibinfo {year} {2020})}\BibitemShut {NoStop}%
\bibitem [{\citenamefont {Echim}\ \emph {et~al.}(2009)\citenamefont {Echim}, \citenamefont {Maggiolo}, \citenamefont {Roth},\ and\ \citenamefont {De~Keyser}}]{echimMagnetosphericGeneratorDriving2009}%
  \BibitemOpen
  \bibfield  {author} {\bibinfo {author} {\bibfnamefont {M.~M.}\ \bibnamefont {Echim}}, \bibinfo {author} {\bibfnamefont {R.}~\bibnamefont {Maggiolo}}, \bibinfo {author} {\bibfnamefont {M.}~\bibnamefont {Roth}},\ and\ \bibinfo {author} {\bibfnamefont {J.}~\bibnamefont {De~Keyser}},\ }\bibfield  {title} {\bibinfo {title} {A magnetospheric generator driving ion and electron acceleration and electric currents in a discrete auroral arc observed by {{Cluster}} and {{DMSP}}},\ }\bibfield  {journal} {\bibinfo  {journal} {Geophysical Research Letters}\ }\textbf {\bibinfo {volume} {36}},\ \href {https://doi.org/10.1029/2009GL038343} {10.1029/2009GL038343} (\bibinfo {year} {2009})\BibitemShut {NoStop}%
\bibitem [{\citenamefont {Imajo}\ \emph {et~al.}(2021)\citenamefont {Imajo}, \citenamefont {Miyoshi}, \citenamefont {Kazama}, \citenamefont {Asamura}, \citenamefont {Shinohara}, \citenamefont {Shiokawa}, \citenamefont {Kasahara}, \citenamefont {Kasaba}, \citenamefont {Matsuoka}, \citenamefont {Wang}, \citenamefont {Tam}, \citenamefont {Chang}, \citenamefont {Wang}, \citenamefont {Angelopoulos}, \citenamefont {Jun}, \citenamefont {Shoji}, \citenamefont {Nakamura}, \citenamefont {Kitahara}, \citenamefont {Teramoto}, \citenamefont {Kurita},\ and\ \citenamefont {Hori}}]{imajo_active_2021}%
  \BibitemOpen
  \bibfield  {author} {\bibinfo {author} {\bibfnamefont {S.}~\bibnamefont {Imajo}}, \bibinfo {author} {\bibfnamefont {Y.}~\bibnamefont {Miyoshi}}, \bibinfo {author} {\bibfnamefont {Y.}~\bibnamefont {Kazama}}, \bibinfo {author} {\bibfnamefont {K.}~\bibnamefont {Asamura}}, \bibinfo {author} {\bibfnamefont {I.}~\bibnamefont {Shinohara}}, \bibinfo {author} {\bibfnamefont {K.}~\bibnamefont {Shiokawa}}, \bibinfo {author} {\bibfnamefont {Y.}~\bibnamefont {Kasahara}}, \bibinfo {author} {\bibfnamefont {Y.}~\bibnamefont {Kasaba}}, \bibinfo {author} {\bibfnamefont {A.}~\bibnamefont {Matsuoka}}, \bibinfo {author} {\bibfnamefont {S.-Y.}\ \bibnamefont {Wang}}, \bibinfo {author} {\bibfnamefont {S.~W.~Y.}\ \bibnamefont {Tam}}, \bibinfo {author} {\bibfnamefont {T.-F.}\ \bibnamefont {Chang}}, \bibinfo {author} {\bibfnamefont {B.-J.}\ \bibnamefont {Wang}}, \bibinfo {author} {\bibfnamefont {V.}~\bibnamefont {Angelopoulos}}, \bibinfo {author} {\bibfnamefont {C.-W.}\ \bibnamefont {Jun}}, \bibinfo {author} {\bibfnamefont
  {M.}~\bibnamefont {Shoji}}, \bibinfo {author} {\bibfnamefont {S.}~\bibnamefont {Nakamura}}, \bibinfo {author} {\bibfnamefont {M.}~\bibnamefont {Kitahara}}, \bibinfo {author} {\bibfnamefont {M.}~\bibnamefont {Teramoto}}, \bibinfo {author} {\bibfnamefont {S.}~\bibnamefont {Kurita}},\ and\ \bibinfo {author} {\bibfnamefont {T.}~\bibnamefont {Hori}},\ }\bibfield  {title} {\bibinfo {title} {Active auroral arc powered by accelerated electrons from very high altitudes},\ }\href {https://doi.org/10.1038/s41598-020-79665-5} {\bibfield  {journal} {\bibinfo  {journal} {Scientific Reports}\ }\textbf {\bibinfo {volume} {11}},\ \bibinfo {pages} {1610} (\bibinfo {year} {2021})}\BibitemShut {NoStop}%
\bibitem [{\citenamefont {{Friis-Christensen}}\ \emph {et~al.}(2006)\citenamefont {{Friis-Christensen}}, \citenamefont {L{\"u}hr},\ and\ \citenamefont {Hulot}}]{friis-christensenSwarmConstellationStudy2006}%
  \BibitemOpen
  \bibfield  {author} {\bibinfo {author} {\bibfnamefont {E.}~\bibnamefont {{Friis-Christensen}}}, \bibinfo {author} {\bibfnamefont {H.}~\bibnamefont {L{\"u}hr}},\ and\ \bibinfo {author} {\bibfnamefont {G.}~\bibnamefont {Hulot}},\ }\bibfield  {title} {\bibinfo {title} {Swarm: {{A}} constellation to study the {{Earth}}'s magnetic field},\ }\href {https://doi.org/10.1186/BF03351933} {\bibfield  {journal} {\bibinfo  {journal} {Earth, Planets and Space}\ }\textbf {\bibinfo {volume} {58}},\ \bibinfo {pages} {BF03351933} (\bibinfo {year} {2006})}\BibitemShut {NoStop}%
\bibitem [{\citenamefont {Knudsen}\ \emph {et~al.}(2017)\citenamefont {Knudsen}, \citenamefont {Burchill}, \citenamefont {Buchert}, \citenamefont {Eriksson}, \citenamefont {Gill}, \citenamefont {Wahlund}, \citenamefont {{\AA}hlen}, \citenamefont {Smith},\ and\ \citenamefont {Moffat}}]{knudsenThermalIonImagers2017}%
  \BibitemOpen
  \bibfield  {author} {\bibinfo {author} {\bibfnamefont {D.~J.}\ \bibnamefont {Knudsen}}, \bibinfo {author} {\bibfnamefont {J.~K.}\ \bibnamefont {Burchill}}, \bibinfo {author} {\bibfnamefont {S.~C.}\ \bibnamefont {Buchert}}, \bibinfo {author} {\bibfnamefont {A.~I.}\ \bibnamefont {Eriksson}}, \bibinfo {author} {\bibfnamefont {R.}~\bibnamefont {Gill}}, \bibinfo {author} {\bibfnamefont {J.-E.}\ \bibnamefont {Wahlund}}, \bibinfo {author} {\bibfnamefont {L.}~\bibnamefont {{\AA}hlen}}, \bibinfo {author} {\bibfnamefont {M.}~\bibnamefont {Smith}},\ and\ \bibinfo {author} {\bibfnamefont {B.}~\bibnamefont {Moffat}},\ }\bibfield  {title} {\bibinfo {title} {Thermal ion imagers and {{Langmuir}} probes in the {{Swarm}} electric field instruments},\ }\href {https://doi.org/10.1002/2016JA022571} {\bibfield  {journal} {\bibinfo  {journal} {Journal of Geophysical Research: Space Physics}\ }\textbf {\bibinfo {volume} {122}},\ \bibinfo {pages} {2016JA022571} (\bibinfo {year} {2017})}\BibitemShut {NoStop}%
\bibitem [{\citenamefont {Yang}\ \emph {et~al.}(2017)\citenamefont {Yang}, \citenamefont {Donovan}, \citenamefont {Liang},\ and\ \citenamefont {Spanswick}}]{yangStatisticalStudyMotion2017}%
  \BibitemOpen
  \bibfield  {author} {\bibinfo {author} {\bibfnamefont {B.}~\bibnamefont {Yang}}, \bibinfo {author} {\bibfnamefont {E.}~\bibnamefont {Donovan}}, \bibinfo {author} {\bibfnamefont {J.}~\bibnamefont {Liang}},\ and\ \bibinfo {author} {\bibfnamefont {E.}~\bibnamefont {Spanswick}},\ }\bibfield  {title} {\bibinfo {title} {A statistical study of the motion of pulsating aurora patches: Using the {{THEMIS All-Sky Imager}}},\ }\href {https://doi.org/10.5194/angeo-35-217-2017} {\bibfield  {journal} {\bibinfo  {journal} {Annales Geophysicae}\ }\textbf {\bibinfo {volume} {35}},\ \bibinfo {pages} {217} (\bibinfo {year} {2017})}\BibitemShut {NoStop}%
\bibitem [{\citenamefont {Oppenheim}\ \emph {et~al.}(2008)\citenamefont {Oppenheim}, \citenamefont {Dimant},\ and\ \citenamefont {Dyrud}}]{oppenheim_large-scale_2008}%
  \BibitemOpen
  \bibfield  {author} {\bibinfo {author} {\bibfnamefont {M.~M.}\ \bibnamefont {Oppenheim}}, \bibinfo {author} {\bibfnamefont {Y.}~\bibnamefont {Dimant}},\ and\ \bibinfo {author} {\bibfnamefont {L.~P.}\ \bibnamefont {Dyrud}},\ }\bibfield  {title} {\bibinfo {title} {Large-scale simulations of 2-{{D}} fully kinetic {{Farley-Buneman}} turbulence},\ }in\ \href@noop {} {\emph {\bibinfo {booktitle} {Annales {{Geophysicae}}}}},\ Vol.~\bibinfo {volume} {26}\ (\bibinfo  {publisher} {Copernicus Publications G{\"o}ttingen, Germany},\ \bibinfo {year} {2008})\ pp.\ \bibinfo {pages} {543--553}\BibitemShut {NoStop}%
\bibitem [{\citenamefont {Oppenheim}\ and\ \citenamefont {Dimant}(2013)}]{oppenheim_kinetic_2013}%
  \BibitemOpen
  \bibfield  {author} {\bibinfo {author} {\bibfnamefont {M.~M.}\ \bibnamefont {Oppenheim}}\ and\ \bibinfo {author} {\bibfnamefont {Y.~S.}\ \bibnamefont {Dimant}},\ }\bibfield  {title} {\bibinfo {title} {Kinetic simulations of 3-{{D Farley}}-{{Buneman}} turbulence and anomalous electron heating},\ }\href {https://doi.org/10.1002/jgra.50196} {\bibfield  {journal} {\bibinfo  {journal} {Journal of Geophysical Research: Space Physics}\ }\textbf {\bibinfo {volume} {118}},\ \bibinfo {pages} {1306} (\bibinfo {year} {2013})}\BibitemShut {NoStop}%
\bibitem [{\citenamefont {Chau}\ and\ \citenamefont {{St.-Maurice}}(2016)}]{chauUnusualRegionFieldaligned2016}%
  \BibitemOpen
  \bibfield  {author} {\bibinfo {author} {\bibfnamefont {J.~L.}\ \bibnamefont {Chau}}\ and\ \bibinfo {author} {\bibfnamefont {J.-P.}\ \bibnamefont {{St.-Maurice}}},\ }\bibfield  {title} {\bibinfo {title} {Unusual 5 m {{E}} region field-aligned irregularities observed from {{Northern Germany}} during the magnetic storm of 17 {{March}} 2015},\ }\href {https://doi.org/10.1002/2016JA023104} {\bibfield  {journal} {\bibinfo  {journal} {Journal of Geophysical Research: Space Physics}\ }\textbf {\bibinfo {volume} {121}},\ \bibinfo {pages} {10,316} (\bibinfo {year} {2016})}\BibitemShut {NoStop}%
\bibitem [{\citenamefont {{St-Maurice}}\ \emph {et~al.}(2023)\citenamefont {{St-Maurice}}, \citenamefont {Huyghebaert}, \citenamefont {Ivarsen},\ and\ \citenamefont {Hussey}}]{st-mauriceNarrowWidthFarleyBuneman2023}%
  \BibitemOpen
  \bibfield  {author} {\bibinfo {author} {\bibfnamefont {J.-P.}\ \bibnamefont {{St-Maurice}}}, \bibinfo {author} {\bibfnamefont {D.}~\bibnamefont {Huyghebaert}}, \bibinfo {author} {\bibfnamefont {M.~F.}\ \bibnamefont {Ivarsen}},\ and\ \bibinfo {author} {\bibfnamefont {G.~C.}\ \bibnamefont {Hussey}},\ }\bibfield  {title} {\bibinfo {title} {Narrow {{Width Farley-Buneman Spectra Above}} 100 km {{Altitude}}},\ }\href {https://doi.org/10.1029/2022JA031191} {\bibfield  {journal} {\bibinfo  {journal} {Journal of Geophysical Research: Space Physics}\ }\textbf {\bibinfo {volume} {128}},\ \bibinfo {pages} {e2022JA031191} (\bibinfo {year} {2023})}\BibitemShut {NoStop}%
\bibitem [{\citenamefont {Opgenoorth}\ \emph {et~al.}(1990)\citenamefont {Opgenoorth}, \citenamefont {H{\"a}gstr{\"o}m}, \citenamefont {Williams},\ and\ \citenamefont {Jones}}]{opgenoorthRegionsStronglyEnhanced1990}%
  \BibitemOpen
  \bibfield  {author} {\bibinfo {author} {\bibfnamefont {H.~J.}\ \bibnamefont {Opgenoorth}}, \bibinfo {author} {\bibfnamefont {I.}~\bibnamefont {H{\"a}gstr{\"o}m}}, \bibinfo {author} {\bibfnamefont {P.~J.~S.}\ \bibnamefont {Williams}},\ and\ \bibinfo {author} {\bibfnamefont {G.~O.~L.}\ \bibnamefont {Jones}},\ }\bibfield  {title} {\bibinfo {title} {Regions of strongly enhanced perpendicular electric fields adjacent to auroral arcs},\ }\href {https://doi.org/10.1016/0021-9169(90)90044-N} {\bibfield  {journal} {\bibinfo  {journal} {Journal of Atmospheric and Terrestrial Physics}\ }\bibinfo {series} {The {{Fourth International EISCAT Workshop}}},\ \textbf {\bibinfo {volume} {52}},\ \bibinfo {pages} {449} (\bibinfo {year} {1990})}\BibitemShut {NoStop}%
\bibitem [{\citenamefont {Lanchester}\ \emph {et~al.}(1996)\citenamefont {Lanchester}, \citenamefont {Kaila},\ and\ \citenamefont {McCrea}}]{lanchester_relationship_1996}%
  \BibitemOpen
  \bibfield  {author} {\bibinfo {author} {\bibfnamefont {B.~S.}\ \bibnamefont {Lanchester}}, \bibinfo {author} {\bibfnamefont {K.}~\bibnamefont {Kaila}},\ and\ \bibinfo {author} {\bibfnamefont {I.~W.}\ \bibnamefont {McCrea}},\ }\bibfield  {title} {\bibinfo {title} {Relationship between large horizontal electric fields and auroral arc elements},\ }\href {https://doi.org/10.1029/95JA02055} {\bibfield  {journal} {\bibinfo  {journal} {Journal of Geophysical Research: Space Physics}\ }\textbf {\bibinfo {volume} {101}},\ \bibinfo {pages} {5075} (\bibinfo {year} {1996})}\BibitemShut {NoStop}%
\bibitem [{\citenamefont {Haerendel}\ \emph {et~al.}(1993)\citenamefont {Haerendel}, \citenamefont {Buchert}, \citenamefont {La~Hoz}, \citenamefont {Raaf},\ and\ \citenamefont {Rieger}}]{haerendel_proper_1993}%
  \BibitemOpen
  \bibfield  {author} {\bibinfo {author} {\bibfnamefont {G.}~\bibnamefont {Haerendel}}, \bibinfo {author} {\bibfnamefont {S.}~\bibnamefont {Buchert}}, \bibinfo {author} {\bibfnamefont {C.}~\bibnamefont {La~Hoz}}, \bibinfo {author} {\bibfnamefont {B.}~\bibnamefont {Raaf}},\ and\ \bibinfo {author} {\bibfnamefont {E.}~\bibnamefont {Rieger}},\ }\bibfield  {title} {\bibinfo {title} {On the proper motion of auroral arcs},\ }\href {https://doi.org/10.1029/92JA02701} {\bibfield  {journal} {\bibinfo  {journal} {Journal of Geophysical Research: Space Physics}\ }\textbf {\bibinfo {volume} {98}},\ \bibinfo {pages} {6087} (\bibinfo {year} {1993})}\BibitemShut {NoStop}%
\bibitem [{\citenamefont {Gjerloev}(2012)}]{gjerloevSuperMAGDataProcessing2012}%
  \BibitemOpen
  \bibfield  {author} {\bibinfo {author} {\bibfnamefont {J.~W.}\ \bibnamefont {Gjerloev}},\ }\bibfield  {title} {\bibinfo {title} {The {{SuperMAG}} data processing technique},\ }\bibfield  {journal} {\bibinfo  {journal} {Journal of Geophysical Research: Space Physics}\ }\textbf {\bibinfo {volume} {117}},\ \href {https://doi.org/10.1029/2012JA017683} {10.1029/2012JA017683} (\bibinfo {year} {2012})\BibitemShut {NoStop}%
\bibitem [{\citenamefont {Sander}\ \emph {et~al.}(1998)\citenamefont {Sander}, \citenamefont {Ester}, \citenamefont {Kriegel},\ and\ \citenamefont {Xu}}]{sander_density-based_1998}%
  \BibitemOpen
  \bibfield  {author} {\bibinfo {author} {\bibfnamefont {J.}~\bibnamefont {Sander}}, \bibinfo {author} {\bibfnamefont {M.}~\bibnamefont {Ester}}, \bibinfo {author} {\bibfnamefont {H.-P.}\ \bibnamefont {Kriegel}},\ and\ \bibinfo {author} {\bibfnamefont {X.}~\bibnamefont {Xu}},\ }\bibfield  {title} {\bibinfo {title} {Density-{{Based Clustering}} in {{Spatial Databases}}: {{The Algorithm GDBSCAN}} and {{Its Applications}}},\ }\href {https://doi.org/10.1023/A:1009745219419} {\bibfield  {journal} {\bibinfo  {journal} {Data Mining and Knowledge Discovery}\ }\textbf {\bibinfo {volume} {2}},\ \bibinfo {pages} {169} (\bibinfo {year} {1998})}\BibitemShut {NoStop}%
\bibitem [{\citenamefont {Camporeale}\ \emph {et~al.}(2018)\citenamefont {Camporeale}, \citenamefont {Wing},\ and\ \citenamefont {Johnson}}]{camporeale_machine_2018}%
  \BibitemOpen
  \bibfield  {author} {\bibinfo {author} {\bibfnamefont {E.}~\bibnamefont {Camporeale}}, \bibinfo {author} {\bibfnamefont {S.}~\bibnamefont {Wing}},\ and\ \bibinfo {author} {\bibfnamefont {J.}~\bibnamefont {Johnson}},\ }\href@noop {} {\emph {\bibinfo {title} {Machine {{Learning Techniques}} for {{Space Weather}}}}}\ (\bibinfo  {publisher} {Elsevier},\ \bibinfo {year} {2018})\BibitemShut {NoStop}%
\bibitem [{\citenamefont {{Monte-Moreno}}\ \emph {et~al.}(2022)\citenamefont {{Monte-Moreno}}, \citenamefont {Yang},\ and\ \citenamefont {{Hern{\'a}ndez-Pajares}}}]{monte-moreno_forecast_2022}%
  \BibitemOpen
  \bibfield  {author} {\bibinfo {author} {\bibfnamefont {E.}~\bibnamefont {{Monte-Moreno}}}, \bibinfo {author} {\bibfnamefont {H.}~\bibnamefont {Yang}},\ and\ \bibinfo {author} {\bibfnamefont {M.}~\bibnamefont {{Hern{\'a}ndez-Pajares}}},\ }\bibfield  {title} {\bibinfo {title} {Forecast of the {{Global TEC}} by {{Nearest Neighbour Technique}}},\ }\href {https://doi.org/10.3390/rs14061361} {\bibfield  {journal} {\bibinfo  {journal} {Remote Sensing}\ }\textbf {\bibinfo {volume} {14}},\ \bibinfo {pages} {1361} (\bibinfo {year} {2022})}\BibitemShut {NoStop}%
\bibitem [{\citenamefont {Valdivia}\ \emph {et~al.}(1996)\citenamefont {Valdivia}, \citenamefont {Sharma},\ and\ \citenamefont {Papadopoulos}}]{valdivia_prediction_1996}%
  \BibitemOpen
  \bibfield  {author} {\bibinfo {author} {\bibfnamefont {J.~A.}\ \bibnamefont {Valdivia}}, \bibinfo {author} {\bibfnamefont {A.~S.}\ \bibnamefont {Sharma}},\ and\ \bibinfo {author} {\bibfnamefont {K.}~\bibnamefont {Papadopoulos}},\ }\bibfield  {title} {\bibinfo {title} {Prediction of magnetic storms by nonlinear models},\ }\href {https://doi.org/10.1029/96GL02828} {\bibfield  {journal} {\bibinfo  {journal} {Geophysical Research Letters}\ }\textbf {\bibinfo {volume} {23}},\ \bibinfo {pages} {2899} (\bibinfo {year} {1996})}\BibitemShut {NoStop}%
\bibitem [{\citenamefont {Antonopoulou}\ \emph {et~al.}(2022)\citenamefont {Antonopoulou}, \citenamefont {Balasis}, \citenamefont {Papadimitriou}, \citenamefont {Boutsi}, \citenamefont {Rontogiannis}, \citenamefont {Koutroumbas}, \citenamefont {Daglis},\ and\ \citenamefont {Giannakis}}]{antonopoulou_convolutional_2022}%
  \BibitemOpen
  \bibfield  {author} {\bibinfo {author} {\bibfnamefont {A.}~\bibnamefont {Antonopoulou}}, \bibinfo {author} {\bibfnamefont {G.}~\bibnamefont {Balasis}}, \bibinfo {author} {\bibfnamefont {C.}~\bibnamefont {Papadimitriou}}, \bibinfo {author} {\bibfnamefont {A.~Z.}\ \bibnamefont {Boutsi}}, \bibinfo {author} {\bibfnamefont {A.}~\bibnamefont {Rontogiannis}}, \bibinfo {author} {\bibfnamefont {K.}~\bibnamefont {Koutroumbas}}, \bibinfo {author} {\bibfnamefont {I.~A.}\ \bibnamefont {Daglis}},\ and\ \bibinfo {author} {\bibfnamefont {O.}~\bibnamefont {Giannakis}},\ }\bibfield  {title} {\bibinfo {title} {Convolutional {{Neural Networks}} for {{Automated ULF Wave Classification}} in {{Swarm Time Series}}},\ }\href {https://doi.org/10.3390/atmos13091488} {\bibfield  {journal} {\bibinfo  {journal} {Atmosphere}\ }\textbf {\bibinfo {volume} {13}},\ \bibinfo {pages} {1488} (\bibinfo {year} {2022})}\BibitemShut {NoStop}%
\end{thebibliography}
\end{document}